\documentclass[prd,preprint,superscriptaddress]{revtex4}
\usepackage{revsymb}
\usepackage{amsmath}
\usepackage{graphicx}
\usepackage{bm}
\newcommand{\be}{\begin{equation}}
\newcommand{\ee}{\end{equation}}
\newcommand{\ben}{\begin{eqnarray}}
\newcommand{\een}{\end{eqnarray}}

\begin{document}
\title{Cosmology with Ricci dark energy}
\author{Sergio del Campo\footnote{E-Mail:
sdelcamp@ucv.cl}} \affiliation{Instituto de F\'{\i}sica,
Pontificia Universidad Cat\'{o}lica de Valpara\'{\i}so, Avenida
Brasil 2950, Casilla 4059, Valpara\'{\i}so, Chile}
\author{J\'{u}lio C. Fabris\footnote{E-mail: fabris@pq.cnpq.br}}
\affiliation{Universidade Federal do Esp\'{\i}rito Santo,
Departamento
de F\'{\i}sica\\
Av. Fernando Ferrari, 514, Campus de Goiabeiras, CEP 29075-910,
Vit\'oria, Esp\'{\i}rito Santo, Brazil}
\author{Ram\'{o}n Herrera\footnote{E-mail: ramon.herrera@ucv.cl}}
\affiliation{Instituto de F\'{\i}sica, Pontificia Universidad
Cat\'{o}lica de Valpara\'{\i}so, Avenida Brasil 2950, Casilla
4059, Valpara\'{\i}so, Chile}
\author{Winfried Zimdahl\footnote{E-mail: winfried.zimdahl@pq.cnpq.br}}
\affiliation{Universidade Federal do Esp\'{\i}rito Santo,
Departamento
de F\'{\i}sica\\
Av. Fernando Ferrari, 514, Campus de Goiabeiras, CEP 29075-910,
Vit\'oria, Esp\'{\i}rito Santo, Brazil}

\begin{abstract}
We assume the cosmological dark sector to consist of pressureless matter and holographic dark energy with a cutoff length proportional to the Ricci scale. The requirement of separate energy-momentum conservation of the components is shown to establish a relation between the matter fraction and the (necessarily time-dependent) equation-of-state parameter of the dark energy.
Focusing on intrinsically adiabatic  pressure perturbations of the dark-energy component, the matter perturbations are found as linear combinations of the total energy-density perturbations
of the cosmic medium and the relative (nonadiabatic) perturbations of the components.
The resulting background dynamics is consistent with observations from supernovae of type Ia, baryonic acoustic oscillations and the differential age of old objects.
The perturbation dynamics, on the other hand, is plagued by instabilities which excludes any phantom-type equation of state. The only stable configuration is singled out by a fixed relation between the present matter fraction $\Omega_{m0}$ and the
present value $\omega_{0}$ of the equation-of-state parameter of the dark energy.
However, this instability-avoiding configuration is only marginally consistent with the observationally  preferred background values of the mentioned parameters.
\end{abstract}
\date{\today}
\maketitle

\section{Introduction}

Despite the many efforts to consistently explain the results of the observations of supernovae of type Ia
(SNIa) in \cite{SN}, the physical nature of the dark sector of the Universe, assumed to consist of dark matter (DM) and dark energy (DE), remains largely mysterious.
The favored cosmological model is the $\Lambda$-cold-dark-matter
($\Lambda$CDM) model which also serves as a reference
for alternative approaches to the DE problem. \textit{Grosso modo}, the $\Lambda$CDM model does well in fitting
most observational data (see, e.g., the recent WMAP 9 results \cite{wmap9}). Nevertheless, there is an ongoing interest in alternative models within General Relativity (GR) itself and beyond it. Although no serious contender seems to be around at the present time, these efforts continue to make sense not only because of the notorious cosmological constant and coincidence problems  (see, e.g. \cite{problems}) but also to test as many potential deviations from the ``standard" description as possible in order to constrain additional parameter sets which are usually introduced in alternative approaches in order to quantify these deviations. Typically, these approaches  ``dynamize" the cosmological constant in terms of scalar fields or fluids with a generally time-dependent equation of state (EoS). Constraining a potential time dependence by the data, e.g., is then crucial for a comparison with the $\Lambda$CDM model.
Holographic models represent a specific class of dynamic approaches to the DE problem.
These models are characterized by a relation between an ultraviolet cutoff and an infrared cutoff \cite{cohen,li,Hsu}. Such relation guarantees that the energy in a given volume does not exceed the energy of a black hole of the same size. The infrared cutoff has to be a cosmological length scale.
For the most obvious choice, the Hubble radius, only models in which dark matter and dark energy are interacting with each other also nongravitationally, give rise to a suitable dynamics \cite{DW,HDE}. Following \cite{li},
there has been a considerable number of investigations based on the future event horizon as cutoff scale \cite{futureEH}. However, all models with a cutoff at the future event horizon suffer from the serious drawback that they cannot describe a transition from decelerated to accelerated expansion. A future event horizon does not exist during the period of decelerated expansion.
A further option that has received attention more recently and which will be the subject of the present paper is a model based on a cutoff length proportional to the Ricci scale.
The role of a distance proportional to the Ricci scale as a causal connection scale for
perturbations was noticed in \cite{brustein}. As a cutoff length in DE
models it was first used in \cite{gao}.
Afterwards, investigations along this line have been carried out in \cite{cai}.
In \cite{xuliluchang,xuwang,xinzhang,rong} observational constraints were obtained on the basis of which Ricci DE was compared with the $\Lambda$CDM model.
The dynamics of perturbations was considered in \cite{fengli,karwan,yuting}.
A number of studies performed in \cite{yi,zhang3,ivdi,luis11,tian-fu,jamil1,zhenhui,luis,jamil2,luis13} include interactions between DM
and Ricci DE. A relation to quantum field theory has been claimed in \cite{broda}.
Various generalizations rely on the cutoff introduced in \cite{granda}.
Other previous work in the field includes \cite{statef,fengzhang,xululi,granda2,suwa,feiyu,yuzhang}.

In the present paper we reconsider the dynamics of a two-component system of pressureless DM and Ricci-type DE
both in the homogeneous and isotropic background and on the perturbative level. While most dynamic DE scenarios start with an assumption for the EoS parameter for the DE, the starting point of holographic models is an expression for the DE energy density from which the EoS is then derived. Moreover, as was pointed out in \cite{SRJW}, the mere definition
 of the holographic DE density generally implies an interaction with the DM component.
 Requiring this interaction to vanish imposes an additional condition on the dynamics.
 Noninteracting Ricci-type DE, in particular, is characterized by a simple relation between the matter fraction and the necessarily time-dependent EoS parameter.
 Therefore it is not compatible with a cosmological constant.
 We shall confront the resulting background dynamics with recent SNIa data, results from baryonic acoustic oscillations (BAO) and from the history $H(z)$ of the Hubble parameter.
 A crucial issue  for Ricci-type DE is the perturbation dynamics \cite{karwan}.
 Based on a gauge-invariant analysis, the matter perturbations are found as a combination of the total and
 the relative energy-density perturbations.
In general, the perturbation dynamics suffers from instabilities. For a phantom-type EoS these should have occurred already before the present time. Consequently, phantom DE is not consistent with our approach.
For present EoS parameters $\omega_{0} > -1$ one obtains growth-rate oscillations and instabilities as well, this time at finite future values of the scale factor $a$. We shall show that there exists just one situation without instabilities at finite values of $a$. It is characterized by a DE saturation parameter already obtained in \cite{karwan}.
We show that for this configuration to be realized, a certain relationship between the current matter content $\Omega_{m0}$  of the Universe and the EoS parameter $\omega_{0}$ is required. Remarkably, under the corresponding condition the
pressure perturbations vanish and the mentioned (unobserved) oscillation disappear.
Moreover, the cosmic coincidence problem is substantially alleviated since holographic Ricci DE itself behaves as nonrelativistic matter at high redshift. There remain, however, tensions between the observationally favored values of $\Omega_{m0}$  and $\omega_{0}$ and the values that are necessary to avoid
instabilities of the perturbation dynamics.

 The structure of the paper is as follows. In Sec. \ref{basic} we recall basic relations for holographic models of DE. The resulting homogeneous and isotropic background dynamics is confronted with observational data in Sec. \ref{observations}. Sec. \ref{general} provides us with the general two-component dynamics of the cosmic medium. The first-order perturbation theory of the model is presented in Secs. \ref{conservation},   \ref{total} and \ref{combining}. On this basis, the final set of coupled equations for the nonadiabatic perturbation dynamics is found in Sec. \ref{nonadiabatic}.
In Sec. \ref{stab} we consider issues of stability and single out a model which is stable for any finite value of the scale factor.
 A summary of the paper is given in the final Sec. \ref{conclusions}.

\section{Background dynamics for Ricci dark energy}
\label{basic}

We start by recalling basic features of holographic DE models in a homogeneous and isotropic background \cite{SRJW}.
The cosmic medium is assumed to be describable  by pressureless DM with energy density
$\rho_{m}$ and a holographic DE component with energy density $\rho_{H}$.
In the spatially flat case Friedmann's equation is
\begin{equation}
3 H^2 = 8\pi\, G (\rho_{m} + \rho_{H})  \  .\label{Fried1}
\end{equation}
In general, both components are not necessarily conserved separately but obey the balance equations
\begin{equation}
\dot{\rho}_{m} + 3 H \rho_{m}  = Q\ ,\qquad
\dot{\rho}_{H} + 3 H (1 + \omega)\rho_{H} = -Q \, , \label{cons2}
\end{equation}
such that the total energy $\rho = \rho_{m} + \rho_{H}$ is conserved.
Here,
$\omega \equiv\frac{p_H}{\rho_{H}} = \frac{p}{\rho_{H}}$ is the equation-of-state (EoS) parameter of the DE and
$p_H$ is the pressure associated with the holographic component.
The acceleration equation can be written
\begin{equation}
\dot{H} = - \frac{3}{2}H^{2}\left(1 + \frac{\omega}{1+r}\right)  \ , \quad \Rightarrow \quad
\frac{d\ln H}{d\ln a}  = - \frac{3}{2}\left(1 + \frac{\omega(a)}{1 + r(a)}\right) \ ,
\label{dH}
\end{equation}
where $r\equiv\frac{\rho_{m}}{\rho_{H}}$ is the ratio of the energy densities.
The \textit{total} effective EoS of the cosmic medium is
\begin{equation}
\frac{p}{\rho} = \frac{\omega}{1 + r}  \ . \label{w}
\end{equation}
According to the balance equations (\ref{cons2}), the ratio $r$ changes as
\begin{equation}
\dot{r} = 3Hr\,\left(1 + r\right)\,\left[\frac{\omega}{1+r} + \frac{Q}{3H\rho_{m}}\right]\ . \label{dr2}
\end{equation}
Following \cite{cohen,li}, we write the holographic energy density as
\begin{equation}
\rho_H=\frac{3\,c^2\,M_p^2}{L^{2}} \ .\label{ans}
\end{equation}
The quantity $L$ is the infrared (IR) cutoff scale and
$M_p=1/\sqrt{8\pi\,G}$ is the reduced Planck mass. The numerical constant $c^{2}$ determines the degree
of saturation of the condition
\begin{equation}
L^{3}\, \rho_H\leq M_{Pl}^{2}\, L \, ,    \label{DEineq}
\end{equation}
which is crucial for any holographic DE model. It states that the energy in a box of size $L$ should not exceed the energy of a black hole of the same size \cite{cohen}.

Differentiation of the expression (\ref{ans}) and use of the energy balances (\ref{cons2}) yields
\begin{equation}
\frac{Q}{\rho_{H}} = 2\frac{\dot{L}}{L} - 3H\left(1+\omega\right) \ . \label{QL}
\end{equation}
In general, there is no reason for $Q$ to vanish. Assuming $Q=0$ provides us with a specific relationship between $\omega$ and the ratio of the rates $\frac{\dot{L}}{L}$ and $H$. Any nonvanishing $Q$ will modify this relationship.

With $Q$ from (\ref{QL}), the general dynamics (\ref{dr2}) of the energy density ratio $r$ becomes
\begin{equation}
\dot{r} = - 3H\,\left(1 + r\right)\,\left[1 + \frac{\omega}{1+r} - \frac{2}{3} \frac{\dot{L}}{H L}\right]\ . \label{drL}
\end{equation}
The case without interaction is characterized by [cf. Eq.~(\ref{dr2})]
\begin{equation}
Q = 0 \quad \Rightarrow\quad \dot{r} = r\left(2\frac{\dot{L}}{L} - 3 H\right)
= 3 H\,r\,\omega \label{QL0}
\end{equation}
with a generally time-dependent $\omega$.
Different choices of the cutoff scale $L$ give rise to different
expressions for the total effective EoS parameter
in (\ref{w}) and to different relations between $\omega$ and $r$. Our interest in the present paper will be the Ricci-scale cutoff.
The role of a distance proportional to the Ricci scale as a causal connection scale for
perturbations was noticed in \cite{brustein}. In \cite{gao} it was used for the first time as a DE cutoff scale.
The Ricci scalar
is $R = 6\left(2H^{2} + \dot{H}\right)$.
For the corresponding cutoff scale one has $L^{2} = 6/R$, i.e.,
\begin{equation}
\rho_H= 3\,c^2\,M_p^2 \,\frac{R}{6} = \alpha\left(2H^{2} + \dot{H}\right) \ ,  \label{rhR}
\end{equation}
where $\alpha = \frac{3c^{2}}{8\pi G}$.
Upon using (\ref{dH})
we obtain
\begin{equation}
\rho_H= \frac{\alpha}{2}\,H^{2}\left(1 - 3\frac{\omega}{1+r}\right)  \label{rh2}
\end{equation}
for the holographic DE density.
Notice that the (not yet known) EoS parameter explicitly enters $\rho_H$. Use of Friedmann's equation provides us with
\begin{equation}
1 =  \frac{c^{2}}{2}\left(1 + r - 3 \omega\right) \quad \Rightarrow\quad \omega = \frac{1}{3}\left(1 + r\right) - \frac{2}{3c^{2}}\ ,
\label{1=}
\end{equation}
which coincides with the result in \cite{karwan}.
Obviously, a constant value of $\omega$ necessarily implies a constant $r$ and vice versa.
The time derivatives of $\omega$ and $r$ are related by $\dot{r} = 3\dot{\omega}$.
The second relation in (\ref{1=}) can be used to express $c^{2}$ in terms of the present values  (subindex $0$) of $\omega$ and $r$:
\begin{equation}
\frac{2}{c^{2}} =  1 + r_{0} - 3 \omega_{0}\ ,\quad \Rightarrow \quad  r = r_{0} + 3(\omega - \omega_{0})\ .
\label{c=}
\end{equation}
The parameter $c$ is related both to $r_{0}$ and $\omega_{0}$.
In a next step we differentiate $\rho_H$ in (\ref{rhR}) which yields
\begin{equation}
\dot{\rho}_H= \alpha \left(4H\dot{H} + \ddot{H}\right)\ . \label{drhoR}
\end{equation}
With the help of (\ref{dH}) and the definition (\ref{rhR}) we derive \cite{SRJW}
\begin{equation}
\dot{\rho}_H + 3H\left(1 + \omega\right)\rho_{H} = - Q
\ , \label{lhs}
\end{equation}
where
\begin{equation}
Q  = - \frac{3H}{1+r}\left[r \omega - \frac{\dot{\omega}}{H}\right]\rho_{H}
\ . \label{Q=}
\end{equation}
Relation (\ref{lhs}) with (\ref{Q=}), which implies that in the general case one has $Q\neq 0$, i.e., both dark components do interact with each other also nongravitationally, is a direct consequence of the ansatz (\ref{rhR}).
The DE balance in Eq.~(\ref{cons2}) may then also be written as $\dot{\rho}_H + 3H\left(1 + \omega_{eff}\right)\rho_{H} = 0$
with an effective EoS parameter
\begin{equation}
\omega_{eff} = \frac{1}{1+r}\left(\omega + \frac{\dot{\omega}}{H}\right) = \frac{\omega + \frac{\dot{\omega}}{H}}{1+r_{0} + 3\left(\omega - \omega_{0}\right)}
\ . \label{wef}
\end{equation}
The present ratio $r_{0}$ is related to the present matter fraction $\Omega_{m0}$ of the Universe by $r_{0} = \frac{\Omega_{m0}}{1 - \Omega_{m0}}$.

According to relation (\ref{Q=}), a constant EoS parameter $\omega$ is compatible with $Q=0$ only for
$\omega=0$, i.e., if $\rho_{H}$ behaves as dust.
If we admit $\dot{\omega}\neq 0$, however, there exists a non trivial case $Q=0$:
\begin{equation}
Q = 0 \quad \Rightarrow\quad r \omega = \frac{\dot{\omega}}{H}
\quad \Rightarrow\quad r = \frac{d\ln \omega}{d\ln a}
\ . \label{rQ0}
\end{equation}
It is this configuration that we shall investigate in the present paper.
Equation (\ref{rQ0}) together with the second relation of (\ref{c=}) is a differential equation for $\omega$ which has the solution
\begin{equation}\label{wsol}
\omega = \omega_{0}\frac{r_{0} - 3\omega_{0}}{r_{0}a^{-(r_{0} - 3\omega_{0})}- 3 \omega_{0}}\ \quad \Rightarrow\quad 1 + \omega =
\frac{r_{0}- \omega_{0}\left(3 - \left(r_{0} - 3\omega_{0}\right)\right)a^{r_{0} - 3\omega_{0}}}
{r_{0} - 3 \omega_{0}a^{(r_{0} - 3\omega_{0})}}\ ,
\end{equation}
where we have normalized the present value of the cosmic scale factor $a$ to $a_{0} = 1$.
The expression for $1 + \omega$ is included here for later reference.
There is no freedom left to choose the equation of state.
It is fixed by the choice of $\rho_{H}$ together with the requirement $Q=0$.
Notice that this is different from the more familiar procedure to deal with (nonholographic) DE, where one starts with an assumption for the EoS parameter and afterwards finds an expression for the DE density by integrating the corresponding balance equation. Here, the starting point is the energy density and the
EoS parameter has to be derived.

Knowing the EoS parameter (\ref{wsol}), it follows from (\ref{c=}) that
\begin{equation}\label{r}
r = r_{0} \frac{r_{0}-3\omega_{0}}{r_{0}-3\omega_{0}a^{\left(r_{0}-3\omega_{0}\right)}}\ ,
\end{equation}
i.e., $r(a)$ is fixed as well.
At high redshifts we have
\begin{equation}
\omega \rightarrow 0 \ ,\qquad r \rightarrow r_0 - 3\omega_0 \ , \qquad\qquad (a \ll 1)
\ . \label{asmall}
\end{equation}
The property that noninteracting Ricci-DE behaves as dust at high redshift was already pointed out in \cite{gao}.
The values in the far-future limit are
\begin{equation}
\omega \rightarrow \omega_0 - \frac{1}{3}r_{0}\ ,\qquad r \rightarrow 0\ , \qquad\qquad (a \gg 1)
\ . \label{alarge}
\end{equation}
The limits in (\ref{asmall}) imply that this model naturally reproduces an early matter-dominated era. For $r_{0} \approx \frac{1}{3}$ and $\omega_{0} \approx -1$, the ratio $r$ approaches $r\approx \frac{10}{3}$ for $a \ll 1$. This value is only roughly ten times larger than the present value $r_{0}$.
For the $\Lambda$CDM model the corresponding difference is about nine orders of magnitude. In this sense, the coincidence problem is considerably alleviated for the the present model. On the other hand, in the opposite limit
$a \gg 1$ the ratio $r$ approaches zero as for the $\Lambda$CDM model. Apparently, the far-future EoS can be of the phantom type for $\omega_0 - \frac{1}{3}r_{0} < -1$.
However, as we shall demonstrate below, such configuration is unstable and does not represent a realistic scenario.

The Hubble rate of our model turns out to be
\begin{equation}
\frac{H}{H_{0}} = a^{-3/2}\,\sqrt{\frac{3\omega_{0}a^{\left(r_{0}-3\omega_{0}\right)}- r_{0}\left[1+r_{0}-3\omega_{0}\right]}
{3\omega_{0}- r_{0}\left[1+r_{0}-3\omega_{0}\right]}}
\ . \label{solHQ0}
\end{equation}
For $a\ll 1$ we recover the Einstein-de Sitter behavior $H \propto a^{-3/2}$.
The total effective EoS is
\begin{equation}
\frac{p}{\rho} = \frac{\omega}{1+r} = \omega_{0}\,\frac{r_{0}-3\omega_{0}}{r_{0}a^{-\left(r_{0}-3\omega_{0}\right)}\left[1 + r_{0}-3\omega_{0}\right] - 3\omega_{0}}
\ . \label{solwtot}
\end{equation}
For the adiabatic sound speed of the DE component we find
\begin{equation}\label{cH}
\frac{\dot{p}_{H}}{\dot{\rho}_{H}} = \omega \left(1-\frac{1}{3}\frac{r}{1+\omega}\right)\
\end{equation}
and the corresponding quantity of the total cosmic medium is
\begin{equation}\label{ctot}
\frac{\dot{p}}{\dot{\rho}} = \frac{\dot{p}_{H}}{\dot{\rho}}
= \omega\frac{1-\frac{1}{3}\frac{r}{1+\omega}}{1+\frac{r}{1+\omega}}\ .
\end{equation}
With the solutions (\ref{wsol}) and (\ref{r}) all these quantities are explicitly known, i.e., the background dynamics is completely solved analytically.
In the following section we perform an actualized confrontation of the background dynamics
with recent observational data.

\section{Observational analysis}
\label{observations}

Our observational analysis of the background dynamics uses the following three tests: the differential age of old objects based on the
$H(z)$ dependence as well as the data from SNIa and from BAO.
A fourth test could potentially be added: the position of the first peak of the anisotropy spectrum of the cosmic microwave background radiation (CMB). However, the CMB test implies integration of the background equations until $z \sim 1.000$ which requires the introduction of the radiative component.
But the inclusion of such radiative component considerably changes the structure of the equations and no analytic expression for $H(z)$ is available. Hence,
we shall limit ourselves to the mentioned three tests for which a reliable estimation is possible.
\par
Based on the evaluation of the age of old galaxies that
have evolved passively \cite{jimenez}, there are 13 observational data available for the differential age \cite{verde,stern,verdebis,ma,mabis}.  Recently, a new set of 21 data has been considered \cite{moresco,ratra}. The basic relation is
\begin{equation}
H(z) = - \frac{1}{1+ z}\frac{dz}{dt}\ .
\end{equation}
The value of the Hubble parameter today can be added to these data, leading to 14 or 22 observational points, depending on the sample used.
\par
The SNIa test is based on the distance modulus $\mu$ which is related to the luminosity distance $D_L$ by
\begin{equation}
\mu =  m - M = 5\log_{10} D_L\ ,\qquad
D_L = \left(1 + z\right)\frac{c}{H_{0}}\int_0^z\frac{dz'}{\sqrt{H\left(z'\right)}}\ .
\end{equation}
In this expression we have restored the velocity of light $c$.
The quantities $m$ and $M$ denote the apparent and the absolute magnitudes, respectively.

Two decisions have to be taken for this test. The first one concerns the choice of the sample. There are many different SNIa data sets, obtained with different techniques.
In some cases, these different samples may give very different results. The second point is the existence of two different calibration methods. One of them uses cosmological relations and
takes into account SNIa with high $z$ (Salt 2), the other one, using astrophysical methods, is suitable for small $z$ (MLCS2k2) \cite{ioav}.
In some cases, the application of different calibrations can lead to different results also. All this makes
the SNIa analysis very delicate. Here, we use the Union 2 sample \cite{union}, calibrated by the Salt 2 method.

Baryonic acoustic oscillations have their origin in oscillations in the photon-baryon plasma at the moment of the decoupling at about $z = 1.090$. They can be characterized by the distance scale \cite{eise},
\begin{equation}
{\cal A} = \frac{\sqrt{\Omega_{m0}}}{[H(z_b)]^{1/3}}\biggr[\frac{1}{z_b}\int_0^{z_b}\frac{dz}{H(z)}\biggl]^{2/3}.
\end{equation}
We shall use the WiggleZ-data \cite{blake} ${\cal A} = 0.474\pm0.034,\ 0.442\pm0.020$ and $0.424\pm0.021$ for the redshifts $z_b = 0.44,\ 0.60$ and $0.73$, respectively.

Generally, the key quantity of a statistical analysis is the
 $\chi^2$ parameter
 \begin{equation}
 \chi^2(x^j) = \sum_i^{n}\frac{(f(x^j)_i^t - f(x^j)_i^o)^2}{\sigma_i},
 \end{equation}
where $f(x^j)_i^t$ is the theoretical evaluation of a given observable, depending on $x^j$ free parameters, $f(x^j)_i^o$ is the corresponding observational value with an error bar $\sigma_i$ and $n$ is the total number of  observational data for the given test. In terms of the $\chi^2$ parameter one defines the
probability distribution function (PDF) by
 \begin{eqnarray}
 P(x^j) = A\,e^{-\chi^2/2},
  \end{eqnarray}
 where $A$ is a normalization constant.
  The estimations for one or for two given parameters are obtained by integrating over the remaining ones.
  For a combination of all tests we use the total $\chi^2$-value $\chi^2_T$,
  \begin{equation}
  \chi^2_T = \chi^2_{H(z)} + \chi^2_{SN} + \chi^2_{BAO}.
  \end{equation}

  Assuming a spatially flat universe, the three free parameters of the model are the density-ratio parameter $r_0 = \frac{\Omega_{mo}}{1 - \Omega_{m0}}$, the EoS $\omega_{0}$ and the reduced
Hubble parameter $h$, defined by $H_{0}= 100\, h\, $km/s/Mpc.

 \begin{center}
\begin{figure}[!t]
\begin{minipage}[t]{0.20\linewidth}
\includegraphics[width=\linewidth]{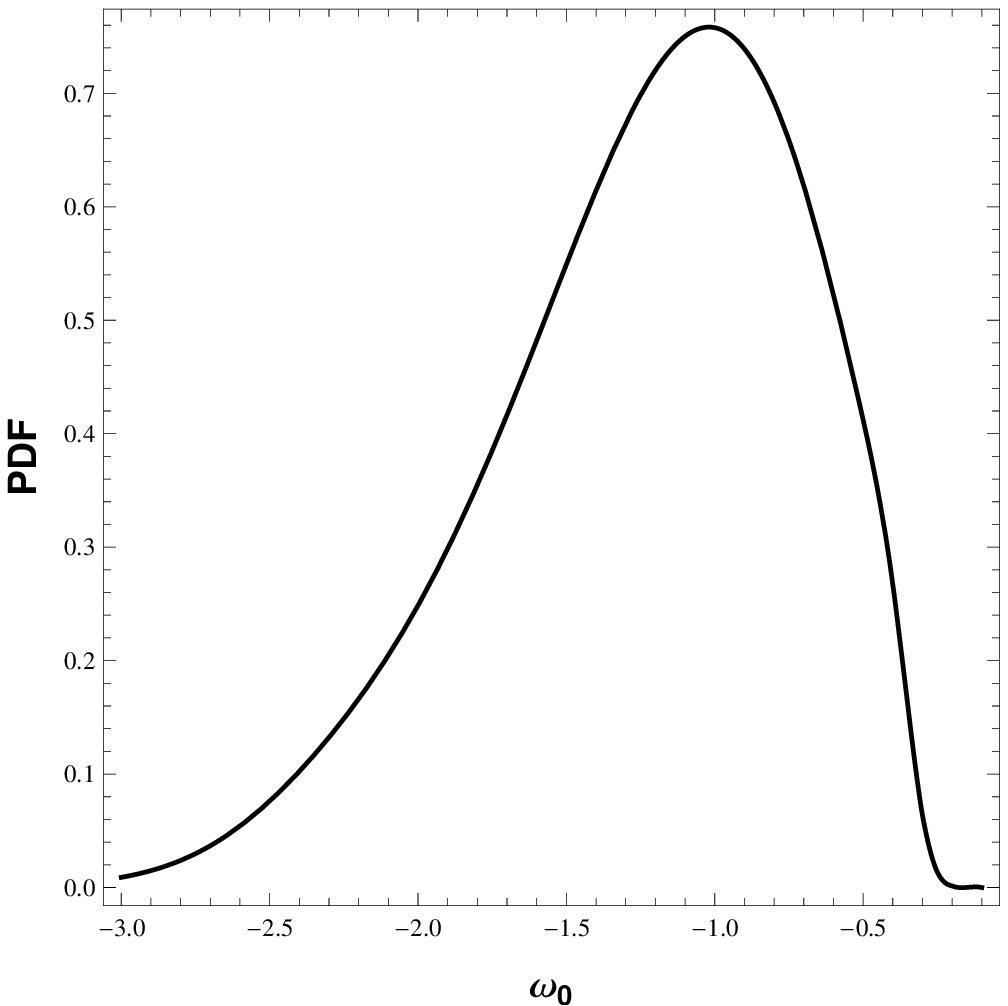}
\end{minipage} \hfill
\begin{minipage}[t]{0.20\linewidth}
\includegraphics[width=\linewidth]{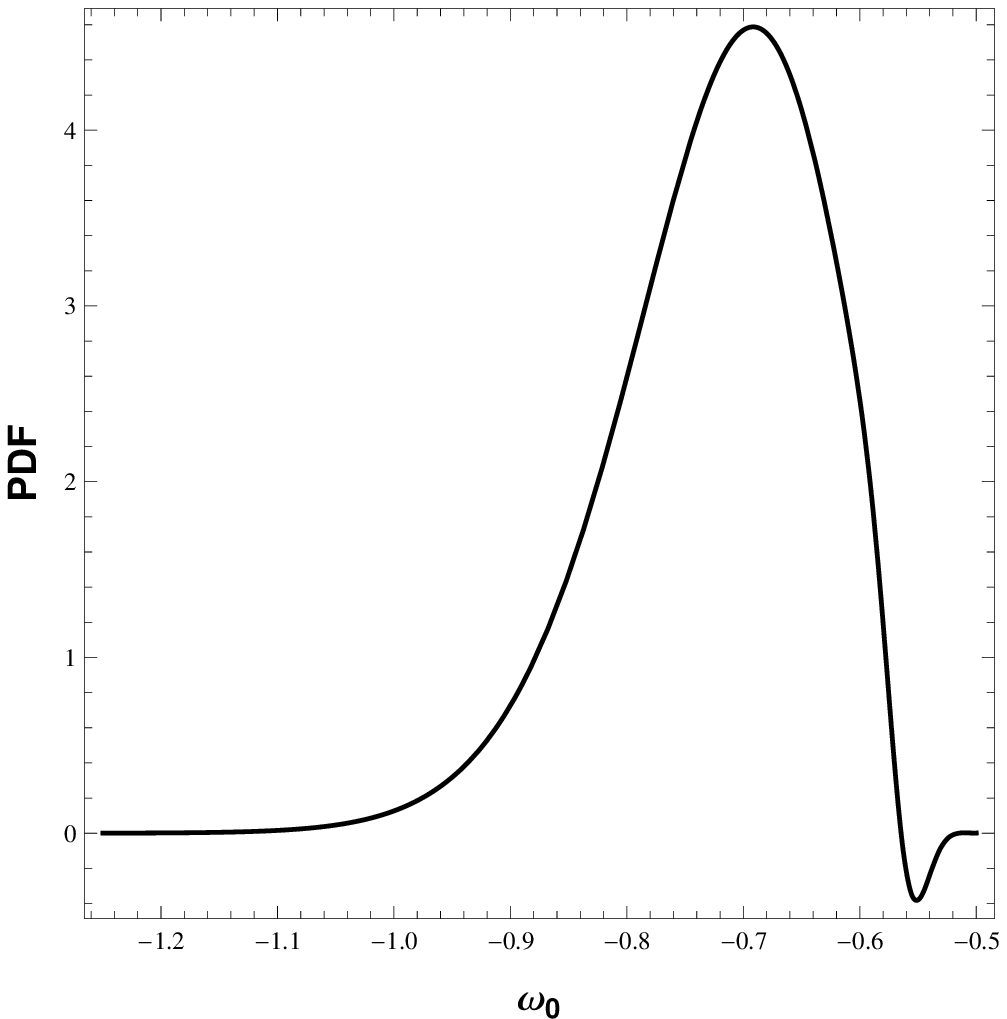}
\end{minipage} \hfill
\begin{minipage}[t]{0.20\linewidth}
\includegraphics[width=\linewidth]{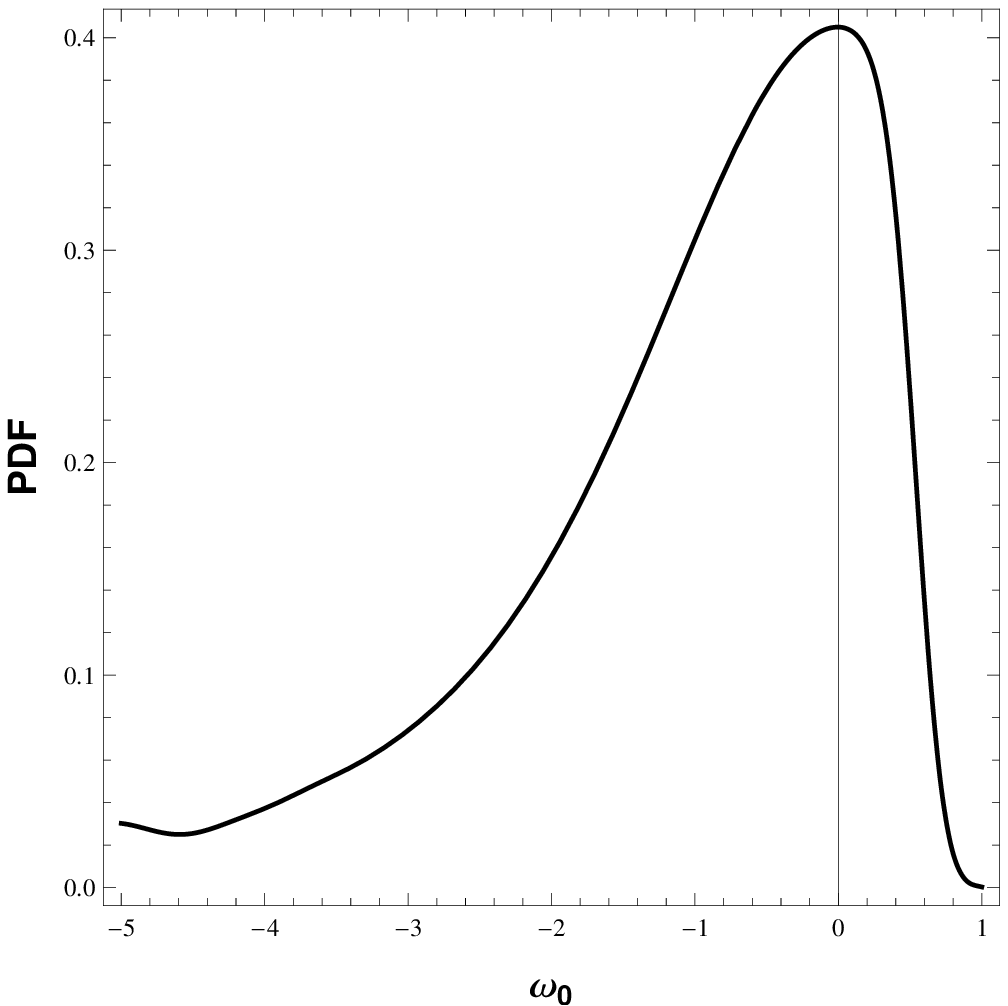}
\end{minipage} \hfill
\begin{minipage}[t]{0.20\linewidth}
\includegraphics[width=\linewidth]{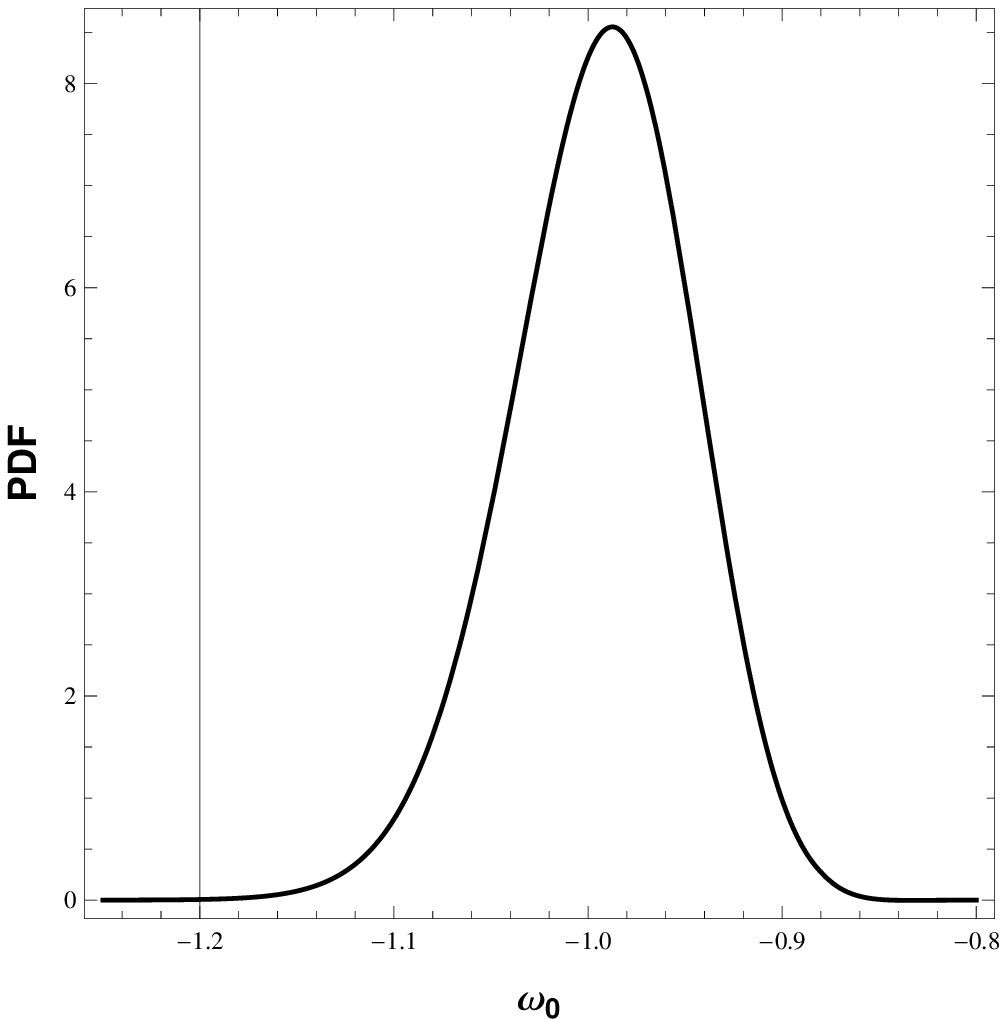}
\end{minipage} \hfill
\begin{minipage}[t]{0.20\linewidth}
\includegraphics[width=\linewidth]{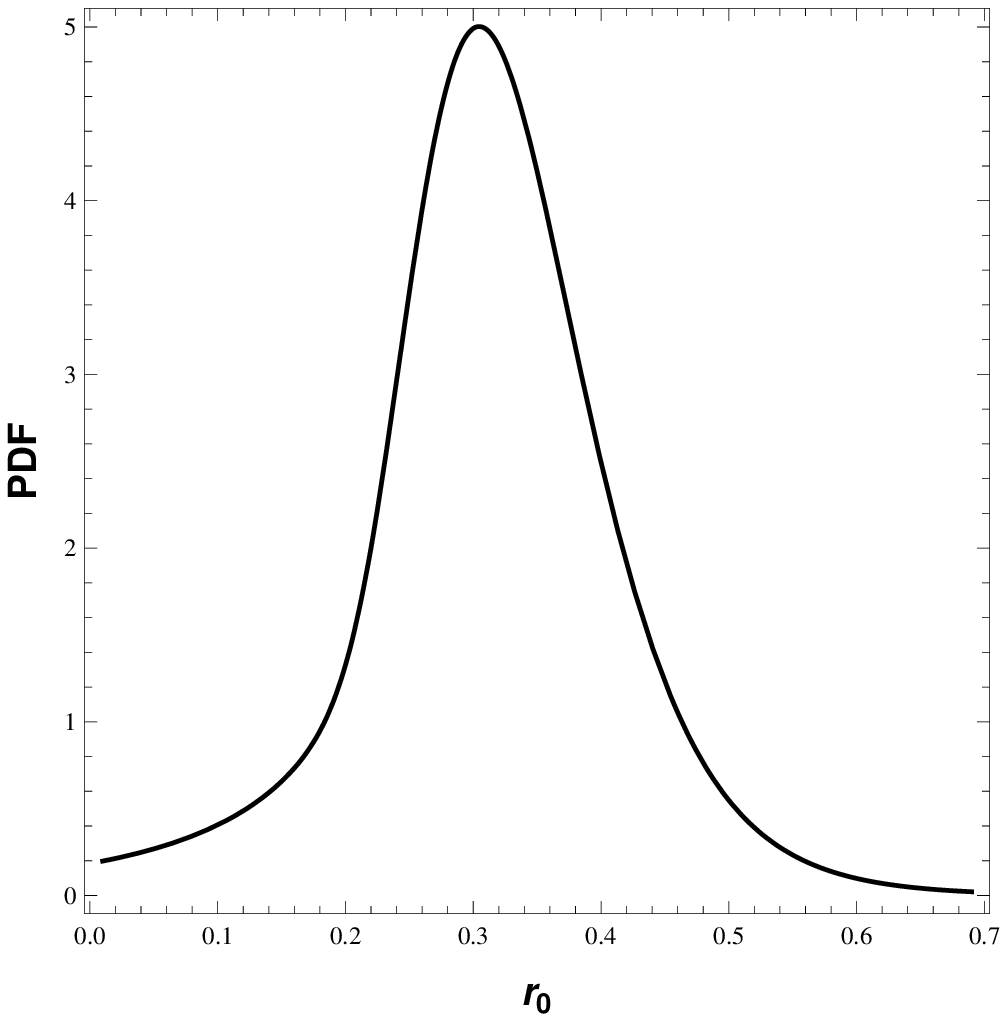}
\end{minipage} \hfill
\begin{minipage}[t]{0.20\linewidth}
\includegraphics[width=\linewidth]{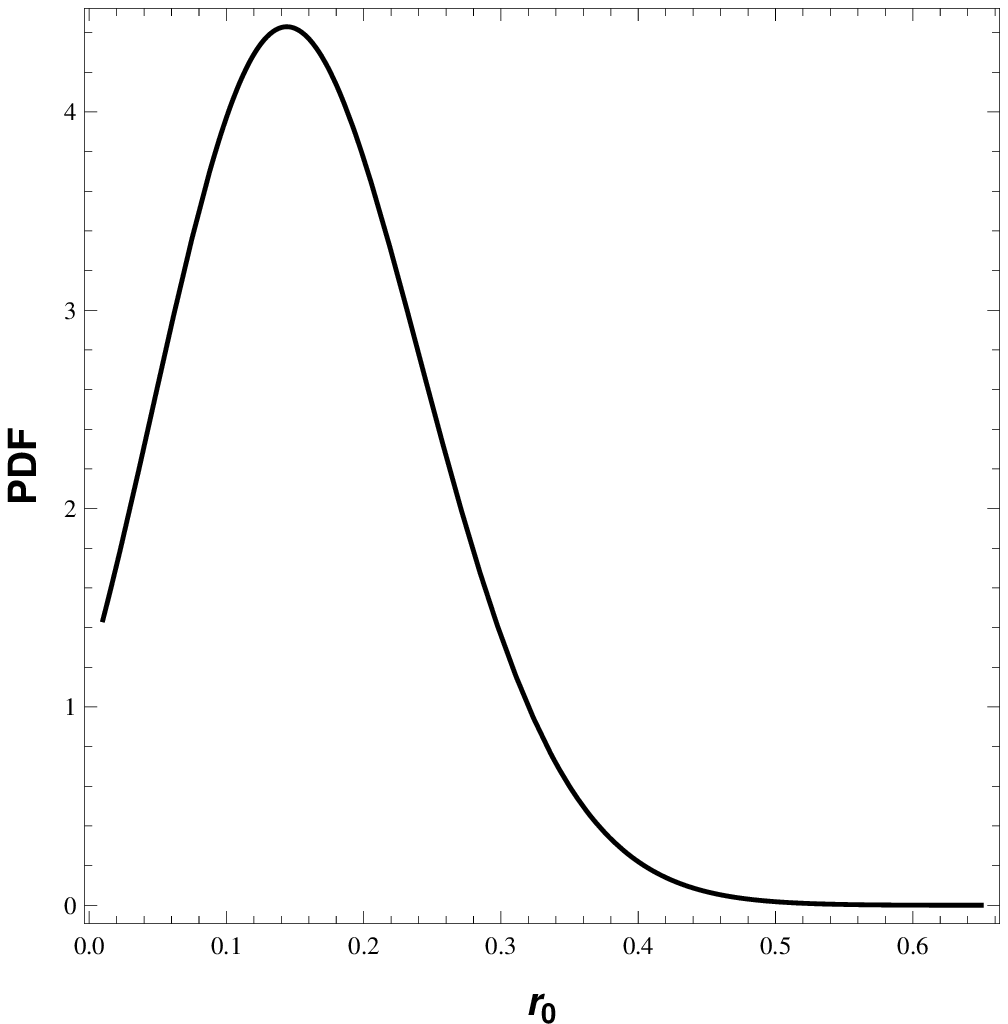}
\end{minipage} \hfill
\begin{minipage}[t]{0.20\linewidth}
\includegraphics[width=\linewidth]{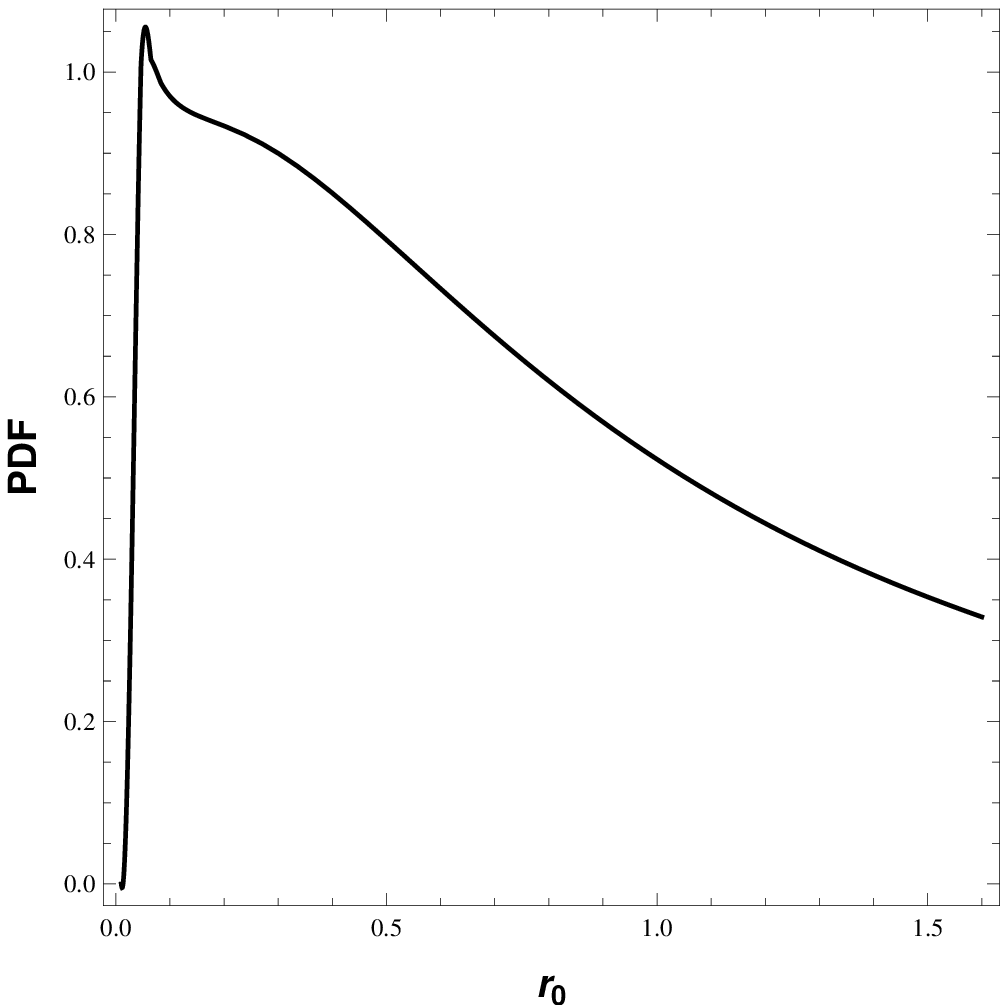}
\end{minipage} \hfill
\begin{minipage}[t]{0.20\linewidth}
\includegraphics[width=\linewidth]{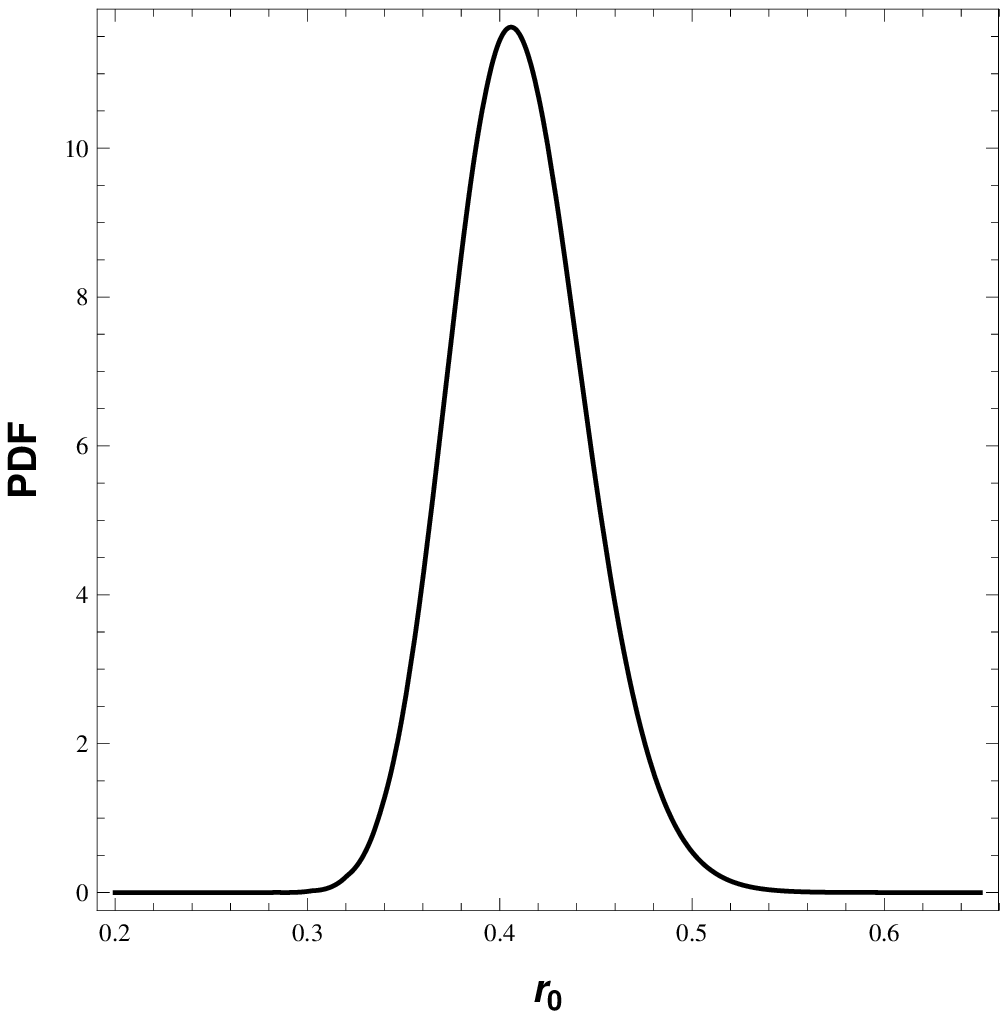}
\end{minipage} \hfill
\caption{One-dimensional PDFs for $\omega_0$ (upper panels) and $r_0$ (lower panels) using, from left to right, $H(z)$, $SNIa$, $BAO$ and the combination of
the three tests.}
\label{fig1}
\end{figure}
\end{center}
\begin{figure}[!t]
\begin{center}
\begin{minipage}[t]{0.50\linewidth}
\includegraphics[width=\linewidth]{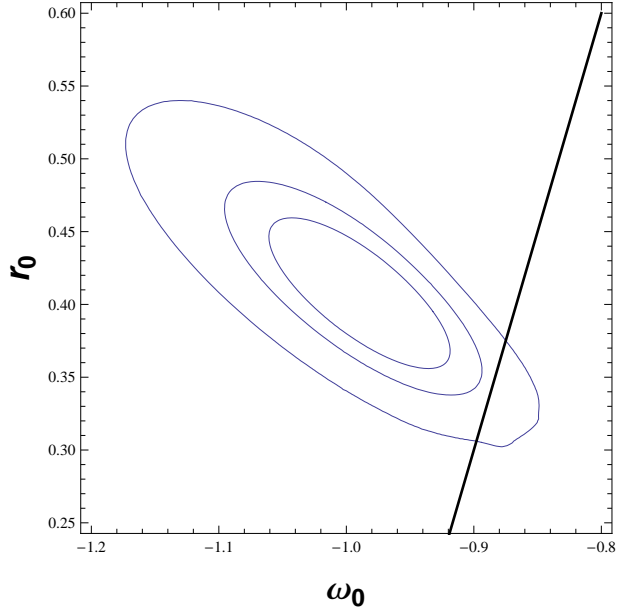}
\end{minipage} \hfill
\caption{Two-dimensional PDF for $\omega_0$ and $r_0$ resulting from a combination of the three tests.
The straight line represents the instability-avoiding configuration $r_{0} = 3\left(1 + \omega_0 \right)$ of Sec. \ref{stab} below.}
\label{fig2}
\end{center}
\end{figure}

In Fig. \ref{fig1} we display the one-dimensional PDFs for each of the tests and for their combination. The results for the density parameter are different for each test.  The combination of
all tests leads to a value of $r_0 \sim 0.4$, corresponding to $\Omega_{m0} \sim 0.29$, roughly in agreement with the $\Lambda$CDM model. For the equation of state parameter we obtain $\omega_0 \sim - 1$, consistent with  the $\Lambda$CDM model as well.
According to the first relation of (\ref{c=}), the parameter $c^{2}$ turns out to be $c^{2} \sim 0.46$.
This value is coincides with the result in \cite{gao}.
 The two-dimensional PDFs at $1\sigma$ ($68\%$ of confidence level), $2\sigma$ ($95\%$ of confidence level) and $3\sigma$ ($99\%$ of confidence level) are shown in Fig. \ref{fig2}.
The estimation for $\omega_0$, based on a combination of the three tests at $2\sigma$, is $\omega_0 = - 0.987^{+0.083}_{-0.100}$, while for $r_0$ we find $r_0 = 0.406^{+0.073}_{-0.061}$.
The straight line represents the combination $r_{0} = 3\left(1 + \omega_0 \right)$ which is singled out by the stability analysis of the perturbation dynamics in Sec. \ref{stab} below. The tension to the results for the background dynamics is obvious, an agreement is possible only at the $3\sigma$ level.

\section{General two-component dynamics}
\label{general}


To study the dynamics of inhomogeneities on the homogeneous and isotropic background of the previous sections
we first consider the general description of the
two-component system. It is characterized by a total energy-momentum tensor
\begin{equation}
T_{ik} = \rho u_{i}u_{k} + p h_{ik}\ , \qquad T_{\ ;k}^{ik} = 0,\
\label{T}
\end{equation}
where $h _{ik}=g_{ik} + u_{i}u_{k}$ and $g_{ik}u^{i}u^{k} = -1$. The quantity $u^{i}$ denotes the total four-velocity of the cosmic substratum. Latin indices run from $0$ to $3$.
The total $T_{ik}$ splits into a matter component (subindex m) and a (holographic) DE component (subindex H),
\begin{equation}\label{Ttot}
T^{ik} = T_{m}^{ik} + T_{H}^{ik},
\end{equation}
with ($A= m, H$)
\begin{equation}\label{TA}
T_{A}^{ik} = \rho_{A} u_A^{i} u^{k}_{A} + p_{A} h_{A}^{ik} \
,\qquad\ h_{A}^{ik} = g^{ik} + u_A^{i} u^{k}_{A} \ .
\end{equation}
For separately conserved fluids we have
\begin{equation}\label{Q}
T_{m\ ;k}^{ik} = 0,\qquad T_{H\ ;k}^{ik} = 0\ .
\end{equation}
Then, the separate energy conservation equations are
\begin{equation}
-u_{mi}T^{ik}_{m\ ;k} = \rho_{m,a}u_{m}^{a} +  \Theta_{m} \rho_{m} = 0\
\label{eb1}
\end{equation}
and
\begin{equation}
-u_{Hi}T^{ik}_{H\ ;k} = \rho_{H,a}u_{H}^{a} +  \Theta_{H} \left(\rho_{H} + p_{H}\right) = 0\ .
\label{eb2}
\end{equation}
In general, each component has its own four-velocity, with $g_{ik}u_{A}^{i}u_{A}^{k} = -1$. The quantities $\Theta_{A}$ are defined as $\Theta_{A} = u^{a}_{A;a}$. For the homogeneous and isotropic background we assume $u_{m}^{a} = u_{H}^{a} = u^{a}$. Likewise, the momentum conservations are written as
\begin{equation}
h_{mi}^{a}T^{ik}_{m\ ;k} = \rho_{m}\dot{u}_{m}^{a} = 0\
\label{mb1}
\end{equation}
and
\begin{equation}
h_{Hi}^{a}T^{ik}_{H\ ;k} = \left(\rho_{H} + p_{H}\right)\dot{u}_{H}^{a} + p_{H,i}h_{H}^{ai} = 0,\
\label{mb2}
\end{equation}
where $\dot{u}_{A}^{a} \equiv u_{A ;b}^{a}u_{A}^{b}$.


Denoting first-order perturbations about the homogeneous and isotropic background by a hat symbol, the perturbed time components of the four-velocities are
\begin{equation}
\hat{u}_{0} = \hat{u}^{0} = \hat{u}_{m}^{0} =\hat{u}_{H}^{0}  = \frac{1}{2}\hat{g}_{00}\ .
\label{u0}
\end{equation}
According to the perfect-fluid structure of both the total energy-momentum tensor (\ref{T}) and the energy-momentum tensors of the components in (\ref{TA}), and with $u_{m}^{a} = u_{H}^{a} = u^{a}$ in the background, we have first-order energy-density perturbations
$\hat{\rho} = \hat{\rho}_{m} + \hat{\rho}_{H}$, pressure perturbations $\hat{p} = \hat{p}_{m} + \hat{p}_{H} = \hat{p}_{H}$
and
\begin{equation}
\hat{T}^{0}_{\alpha} = \hat{T}^{0}_{m\alpha} + \hat{T}^{0}_{H\alpha}\quad\Rightarrow\quad
\left(\rho + p\right)\hat{u}_{\alpha} = \rho_{m}\hat{u}_{m\alpha} + \left(\rho_{H} + p_{H}\right)\hat{u}_{H\alpha}
\ .
\label{T0al}
\end{equation}
In linear order the spatial components of the $4-$ accelerations are
\begin{equation}\label{dotu}
\hat{\dot{u}}_{\mu} = \dot{\hat{u}}_{\mu} - \frac{1}{2}\hat{g}_{00,\mu}, \quad \hat{\dot{u}}_{(m)\mu} = \dot{\hat{u}}_{(m)\mu} - \frac{1}{2}\hat{g}_{00,\mu}, \quad\hat{\dot{u}}_{(H)\mu} = \dot{\hat{u}}_{(H)\mu} - \frac{1}{2}\hat{g}_{00,\mu}\ .
\end{equation}
For the first-order pressure gradient terms we find (recall $p_{H} = p$)
\begin{equation}\label{pressuregrad}
\widehat{p_{,c}h_{H\mu}^{c}} = \hat{p}_{,\mu} + \dot{p}\hat{u}_{H\mu}, \qquad \widehat{p_{,c}h_{\mu}^{c}}
= \hat{p}_{,\mu} + \dot{p}\hat{u}_{\mu}\ .
\end{equation}
From the matter-momentum conservation (\ref{mb1}) it follows that
\begin{equation}\label{mbhat}
\dot{u}_{m}^{a} = 0 \quad \Rightarrow \dot{\hat{u}}_{(m)\mu} - \frac{1}{2}\hat{g}_{00,\mu} = 0
\quad \Rightarrow \dot{\hat{u}}_{(m)\mu} = \frac{1}{2}\hat{g}_{00,\mu}\ .
\end{equation}
According to (\ref{T0al}) the differences between $\hat{u}_{\alpha}$ and the corresponding quantities of the components are
\begin{equation}\label{}
\hat{u}_{\alpha} - \hat{u}_{H\alpha} = \frac{\rho_{m}}{\rho + p}\left(\hat{u}_{m\alpha} - \hat{u}_{H\alpha}\right)\ , \quad
\hat{u}_{\alpha} - \hat{u}_{m\alpha} = \frac{\rho_{H} + p_{H}}{\rho + p}
\left(\hat{u}_{H\alpha} - \hat{u}_{m\alpha}\right)\ .
\end{equation}

Restricting ourselves to scalar perturbations, the perturbed line element can be written
\begin{equation}
\mbox{d}s^{2} = - \left(1 + 2 \phi\right)\mbox{d}t^2 + 2 a^2
F_{,\alpha }\mbox{d}t\mbox{d}x^{\alpha} +
a^2\left[\left(1-2\psi\right)\delta _{\alpha \beta} + 2E_{,\alpha
\beta} \right] \mbox{d}x^\alpha\mbox{d}x^\beta \ .\label{ds}
\end{equation}
Furthermore, we define the (three-) scalar quantities $v$, $v_{m}$ and $v_{H}$ by
\begin{equation}
a^2\hat{u}^{\mu} + a^2F_{,\mu} = \hat{u}_{\mu} \equiv v_{,\mu} \
\label{}
\end{equation}
and
\begin{equation}
a^2\hat{u}^{\mu}_{m} + a^2F_{,\mu} = \hat{u}_{m\mu} \equiv v_{m,\mu} \ ,\quad
a^2\hat{u}^{\mu}_{H} + a^2F_{,\mu} = \hat{u}_{H\mu} \equiv v_{H,\mu}\ ,
\label{}
\end{equation}
respectively.

From the definitions of $\Theta$, $\Theta_{m}$ and $\Theta_{H}$ it follows that
\begin{equation}
\hat{\Theta} = \frac{1}{a^2}\left(\Delta v +\Delta \chi\right) -
3\dot{\psi} - 3 H\phi \ ,
\label{Thetaexpl0}
\end{equation}
where $\Delta$ is the three-dimensional Laplacian,
\begin{equation}
\chi \equiv a^2\left(\dot{E} -F\right) \
\label{}
\end{equation}
and
\begin{equation}
\hat{\Theta}_{m} = \frac{1}{a^2}\left(\Delta v_{m} +\Delta \chi\right) -
3\dot{\psi} - 3 H\phi \ , \quad \hat{\Theta}_{H} = \frac{1}{a^2}\left (\Delta v_{H} +\Delta \chi\right) -
3\dot{\psi} - 3 H\phi \ .
\label{Thetaexpl}
\end{equation}

\section{Conservation equations in linear order}
\label{conservation}

At first order, the energy balances (\ref{eb1}) and (\ref{eb2}) are
\begin{equation}\label{ebm1}
\dot{\hat{\rho}}_{m} + \dot{\rho}_{m}\hat{u}^{0} + \hat{\Theta}_{m}\rho_{m} + \Theta\hat{\rho}_{m} = 0\
\end{equation}
and
\begin{equation}\label{ebH1}
\dot{\hat{\rho}}_{H} + \dot{\rho}_{H}\hat{u}^{0} + \hat{\Theta}_{H}\left(\rho_{H} + p_{H}\right) + \Theta\left(\hat{\rho}_{H} + \hat{p}_{H}\right) = 0\ ,
\end{equation}
respectively. The total first-order energy conservation takes the form
\begin{equation}\label{ebtot1}
\dot{\hat{\rho}} + \dot{\rho}\hat{u}^{0} + \hat{\Theta}\left(\rho + p\right) + \Theta\left(\hat{\rho} + \hat{p}\right) = 0\ .
\end{equation}
Comparing (\ref{ebm1}), (\ref{ebH1}) and (\ref{ebtot1})  one finds
\begin{equation}\label{Thetasum}
\hat{\Theta} = \frac{\rho_{m}}{\rho + p}\hat{\Theta}_{m} + \frac{\rho_{H} + p_{H}}{\rho + p}\hat{\Theta}_{H}\
\end{equation}
with $p = p_{H}$.
The separate momentum conservation equations are given by (\ref{mb1}) and (\ref{mb2}).
Additionally, we have the total momentum conservation
\begin{equation}
\left(\rho_{m} + \rho_{H} + p_{H}\right)\dot{u}^{a} + p_{H,i}h^{ai} = 0\ .
\label{mbperttot}
\end{equation}
For the total energy-density perturbations
$\delta \equiv \frac{\hat{\rho}}{\rho}$ equation (\ref{ebtot1}) yields
\begin{equation}\label{ddelta}
\dot{\delta} -\Theta\frac{p}{\rho}\delta  - \frac{\dot{\rho}}{\rho}\phi + \hat{\Theta}\left(1 + \frac{p}{\rho}\right) + \Theta \frac{\hat{p}}{\rho}= 0\ ,
\end{equation}
where we have used $\hat{u}^{0} = - \phi$.
The momentum balance (\ref{mbperttot}) for the cosmic medium as a whole together with the first relation of (\ref{dotu}) and the second relation of  (\ref{pressuregrad}) provides us with
\begin{equation}\label{mbaltot}
\left(\rho + p\right)\left[\dot{v} + \phi\right] + \hat{p}^{c} = 0\ ,
\end{equation}
where $\hat{p}^{c} = \hat{p} + \dot{p}v$.
In terms of the fractional perturbation $\delta_{m} \equiv \frac{\hat{\rho}_{m}}{\rho_{m}}$, the matter energy conservation (\ref{ebm1}) can be written as
\begin{equation}\label{dDeltam}
\dot{\delta}_{m} - \frac{\dot{\rho}_{m}}{\rho_{m}}\phi + \hat{\Theta}_{m} = 0\ .
\end{equation}
The matter momentum balance (\ref{mb1}) together with the second relation of (\ref{dotu}) results in
\begin{equation}\label{dotvm}
\dot{v}_{m} + \phi = 0\ .
\end{equation}
With $\delta_{H} \equiv \frac{\hat{\rho}_{H}}{\rho_{H}}$ the energy conservation (\ref{ebH1})  for the DE  component is
\begin{equation}\label{ddeltaH}
\dot{\delta}_{H} + \frac{\dot{\rho}_{H}}{\rho_{H}}\delta_{H} - \frac{\dot{\rho}_{H}}{\rho_{H}}\phi + \hat{\Theta}_{H}\left(1 + \frac{p_{H}}{\rho_{H}}\right) + \Theta \left(\delta_{H} + \frac{\hat{p}_{H}}{\rho_{H}}\right)= 0\ .
\end{equation}
In the following it will be convenient to introduce the quantity
\begin{equation}\label{}
D_{H} \equiv \frac{\hat{\rho}_{H}}{\rho_{H} + p_{H}} = \frac{\delta_{H}}{1 + \omega}\ .
\end{equation}
In terms of $D_{H}$
eq.~(\ref{ddeltaH}) then takes the form
\begin{equation}\label{dDH}
\dot{D}_{H} - \Theta\frac{\dot{p}_{H}}{\dot{\rho}_{H}}D_{H} + \Theta\phi
+ \hat{\Theta}_{H} + \Theta\frac{\hat{p}_{H}}{\rho_{H} + p_{H}} = 0\ .
\end{equation}

The dark-energy momentum balance (\ref{mb2}) together with the third relation of (\ref{dotu}) and the first relation of  (\ref{pressuregrad}) result in
\begin{equation}\label{mbalHc}
\left(\rho_{H} + p_{H}\right)\left[\dot{v}_{H} + \phi\right] + \hat{p}^{c_{_{H}}} = 0
\quad \Rightarrow\quad \dot{v}_{H} + \phi = - \frac{\hat{p}^{c_{_{H}}}}{\rho_{H} + p_{H}}
\equiv - P^{c_{_{H}}}\ ,
\end{equation}
where $\hat{p}^{c_{_{H}}} = \hat{p}_{H} + \dot{p}_{H}v_{H} = \hat{p} + \dot{p}v_{H} $.

Our final goal in this paper is to calculate the matter-density perturbations.
To this purpose we shall solve the coupled system of total energy perturbations $\delta$ and relative energy-density perturbations $\delta_{m} - D_{H}$. In the following section
we start by establishing the equation for the total energy-density perturbations.

\section{Total energy-density perturbations}
\label{total}

We consider Eq. (\ref{ddelta}) and introduce therein the gauge-invariant quantities
\begin{equation}\label{}
\delta^{c} = \delta + \frac{\dot{\rho}}{\rho}v\ ,\quad
\hat{\Theta}^{c} = \hat{\Theta}  + \dot{\Theta}v\ .
\end{equation}
Then, Eq.~(\ref{ddelta}) is rewritten as
\begin{equation}\label{ddeltac}
\dot{\delta}^{c} - \Theta\frac{p}{\rho}\delta^{c}
-\frac{\dot{\rho}}{\rho}\left(\dot{v} + \phi\right)
+ \frac{\Theta}{\rho}\hat{p}^{c}
+ \left(1+\frac{p}{\rho}\right)\hat{\Theta}^{c} = 0\ .
\end{equation}
Combination of the energy conservation (\ref{ddeltac}) and the momentum conservation (\ref{mbaltot}) yields
\begin{equation}\label{balcomb}
\dot{\delta}^{c} - \Theta \frac{p}{\rho}\delta^{c} + \left(1 + \frac{p}{\rho}\right)\hat{\Theta}^{c} = 0\ .
\end{equation}
The perturbation $\hat{\Theta}$ has to be determined from the perturbed Raychaudhuri equation for $\Theta$. Neglecting shear and vorticity, the Raychaudhuri equation is
\begin{equation}
\dot{\Theta} + \frac{1}{3}\Theta^{2} - \dot{u}^{a}_{;a} + 4\pi G \left(\rho + 3
p\right) = 0\ .\label{Ray}
\end{equation}
In terms of the gauge-invariant variables one finds, at linear order,
\begin{equation}\label{dThetacfin}
\dot{\hat{\Theta}}^{c} + \frac{2}{3}\Theta\hat{\Theta}^{c} + 4\pi G\rho\delta^{c}
 - \dot{u}^{a}_{:a} = 0\ .
\end{equation}
In a next step we have to differentiate (\ref{balcomb}) and to insert (\ref{dThetacfin}) into the resulting expression. The remaining $\hat{\Theta}^{c}$ terms can be eliminated by (\ref{balcomb}) again.
Using also
\begin{equation}
\dot{u}^m_{;m} =
- \frac{1}{a^2}
\left(\frac{\Delta \hat{p} + \dot{p}\Delta v}{\rho + p}\right)
= - \frac{1}{a^2}\frac{\Delta \hat{p}^{c}}{\rho + p}\ ,
\label{}
\end{equation}
the equation for $\delta^{c}$ becomes
\begin{equation}
\ddot{\delta}^{c} + \left[2-6\frac{p}{\rho}+ 3\frac{\dot{p}}{\dot{\rho}}\right]H\dot{\delta}^{c}
- \left[\frac{3}{2} + 12\frac{p}{\rho} - \frac{9}{2}\frac{p^{2}}{\rho^{2}} - 9\frac{\dot{p}}{\dot{\rho}}\right]H^{2}\delta^{c} - \frac{1}{a^2}\frac{\Delta \hat{p}^{c}}{\rho}= 0 \ .
  \label{}
\end{equation}
Changing to $a$ as independent variable ($\delta^{\prime} \equiv \frac{d \delta^{c}}{d a}$) and transforming into the $k$ space, we arrive at
\begin{equation}
\delta^{c\prime\prime} + \left[\frac{3}{2}-\frac{15}{2}\frac{p}{\rho}+ 3\frac{\dot{p}}{\dot{\rho}}\right]\frac{\delta^{c\prime}}{a}
- \left[\frac{3}{2} + 12\frac{p}{\rho} - \frac{9}{2}\frac{p^{2}}{\rho^{2}} - 9\frac{\dot{p}}{\dot{\rho}}
\right]\frac{\delta^{c}}{a^{2}}
+ \frac{k^{2}}{a^{2}H^{2}}\frac{\hat{p}^{c}}{\rho a^{2}}
= 0\ .
  \label{dddeltak}
\end{equation}
Equation (\ref{dddeltak}) governs the behavior of the total energy-density perturbations. As we shall see,
via the $\hat{p}^{c}$ term the perturbations $\delta^{c}$ are coupled to the relative perturbations  $\delta_{m} - D_{H}$.

 \section{Combining the separate conservation equations}
 \label{combining}

Now we combine the separate energy conservation equations (\ref{dDeltam}) and (\ref{dDH}) of the components and define $S_{mH} \equiv \delta_{m} - D_{H}$. Then
\begin{equation}\label{dSmH}
\dot{S}_{mH} = \Theta\left(P_{H} - \frac{\dot{p}_{H}}{\dot{\rho}_{H}}D_{H}\right) + \hat{\Theta}_{H} - \hat{\Theta}_{m}\ ,
\end{equation}
where $P_{H} = \frac{\hat{p}_{H}}{\rho_{H} + p_{H}}$.
Combining the momentum balances (\ref{dotvm}) and (\ref{mbalHc}) results in
\begin{equation}\label{ddiffv=pc}
\left(v_{H} - v_{m}\right)^{\displaystyle\cdot} = - P^{c_{H}}\ .
\end{equation}
Because of the structure of the first-order expressions in (\ref{Thetaexpl}) one has
\begin{equation}\label{}
\hat{\Theta}_{H} - \hat{\Theta}_{m} = \frac{1}{a^{^{2}}}\Delta\left(v_{H} - v_{m}\right)\ .
\end{equation}
Equation~(\ref{dSmH}) then becomes
\begin{equation}\label{dSmH1}
\dot{S}_{mH} = \Theta\left(P_{H} - \frac{\dot{p}_{H}}{\dot{\rho}_{H}}D_{H}\right) + \frac{1}{a^{^{2}}}\Delta\left(v_{H} - v_{m}\right) .
\end{equation}
Differentiation of (\ref{dSmH1}) yields
\begin{equation}\label{}
\ddot{S}_{mH} = \left[\Theta\left(P_{H} - \frac{\dot{p}_{H}}{\dot{\rho}_{H}}D_{H}\right)\right]^{\displaystyle\cdot} - 2 H \frac{1}{a^{^{2}}}\Delta\left(v_{H} - v_{m}\right)
+ \frac{1}{a^{^{2}}}\Delta\left(v_{H} - v_{m}\right)^{\displaystyle\cdot}\ .
\end{equation}
Using here (\ref{dSmH1}) again and also (\ref{ddiffv=pc}) results in
\begin{equation}\label{ddSmH}
\ddot{S}_{mH} + 2H \dot{S}_{mH} = \left[\Theta\left(P_{H} - \frac{\dot{p}_{H}}{\dot{\rho}_{H}}D_{H}\right)\right]^{\displaystyle\cdot}
+ 2 H \Theta\left(P_{H} - \frac{\dot{p}_{H}}{\dot{\rho}_{H}}D_{H}\right)
- \frac{\Delta P^{c_{H}}}{a^{2}} \ .
\end{equation}
The difference $P_{H} - \frac{\dot{p}_{H}}{\dot{\rho}_{H}}D_{H}$ describes the deviation of the DE pressure perturbations from being adiabatic. It is zero for purely adiabatic DE perturbations.
We discuss this issue in more detail in the following section.

\section{Nonadiabaticity and final set of equations}
\label{nonadiabatic}

Generally, the deviation from adiabaticity in a
two-component system with components $m$ (pressureless) and $H$  is
\begin{eqnarray}
\frac{\hat{p}}{\rho + p} -
\frac{\dot{p}}{\dot{\rho}}\frac{\hat{\rho}}{\rho + p}  = P^{c} - \frac{\dot{p}}{\dot{\rho}}D^{c} &=&
 \frac{\rho_{H} + p_{H}}{\rho + p}
\left(\frac{\hat{p}_{H}}{\rho_{H} + p_{H}} -
\frac{\dot{p}_H}{\dot{\rho}_H}\frac{\hat{\rho}_{H}}{\rho_{H} +
p_{H}}  \right)\nonumber\\
\ \nonumber\\&& + \frac{\rho_{m} \left(\rho_{H} + p_{H}\right)} {\left(\rho +
p\right)^2} \frac{\dot{p}_H}{\dot{\rho}_H}
\left[\frac{\hat{\rho}_{H}}{\rho_{H} + p_{H}} -
\frac{\hat{\rho}_{m}}{\rho_{m}} \right] \ .
\label{nadgen}
\end{eqnarray}
Let us consider the combination
$\hat{p}_{H}-
\frac{\dot{p}_H}{\dot{\rho}_H}\hat{\rho}_{H}$.
With $p_{H} = \omega\rho_{H}$ one has
\begin{equation}\label{hatpH-}
\hat{p}_{H}-
\frac{\dot{p}_H}{\dot{\rho}_H}\hat{\rho}_{H}  = \hat{\omega} \rho_{H} + \left(\omega - \frac{\dot{p}_H}{\dot{\rho}_H}\right)\hat{\rho}_{H} \ ,
\end{equation}
where $\omega$ is given by the solution (\ref{wsol}) and the adiabatic DE sound speed $\frac{\dot{p}_H}{\dot{\rho}_H}$ by (\ref{cH}).
Because of (\ref{cH}) the combination
(\ref{hatpH-}) then results in
\begin{equation}\label{hatpH-1}
\hat{p}_{H}-
\frac{\dot{p}_H}{\dot{\rho}_H}\hat{\rho}_{H}  = \rho_{H}\left[\hat{\omega} + \frac{\omega}{3} r D_{H}\right]\ ,
\end{equation}
which is a gauge-invariant expression. In general, now an assumption for the perturbed EoS parameter $\hat{\omega}$ is necessary to proceed. We shall consider here the
adiabatic case
\begin{equation}\label{holad}
\hat{p}_{H} =
\frac{\dot{p}_H}{\dot{\rho}_H}\hat{\rho}_{H}\qquad
\Rightarrow\qquad \hat{\omega} = - \frac{r \omega}{3}D_{H}\ .
\end{equation}
This assumption of an adiabatic DE component allows us to relate the otherwise undetermined perturbation $\hat{\omega}$ of the EoS parameter
to the DE energy perturbation $D_{H}$.
Under these circumstances Eq. (\ref{ddSmH}) reduces to
\begin{equation}\label{ddSmHad}
\ddot{S}_{mH} + 2H \dot{S}_{mH} =
- \frac{\Delta P^{c_{H}}}{a^{2}} \ .
\end{equation}
We emphasize that the total perturbation dynamics remains nonadiabatic due to the two-component nature of the medium.
With an adiabatic DE component, the general relation (\ref{nadgen}) simplifies to
\begin{equation}
\frac{\hat{p}^{c}}{\rho + p} -
\frac{\dot{p}}{\dot{\rho}}\frac{\hat{\rho}^{c}}{\rho + p} = -
\frac{\rho_{m} \left(\rho_{H} + p_{H}\right)} {\left(\rho +
p\right)^2} \frac{\dot{p}_H}{\dot{\rho}_H}S_{mH} \ ,\label{nad}
\end{equation}
or
\begin{equation}\label{pcnad}
\hat{p}^{c} =  \frac{\dot{p}}{\dot{\rho}} \hat{\rho}^{c}   -
\frac{\rho_{m} \left(\rho_{H} + p_{H}\right)} {\left(\rho +
p\right)} \frac{\dot{p}_H}{\dot{\rho}_H}S_{mH}
\quad \Rightarrow\quad \hat{p}^{c} =  \frac{\dot{p}}{\dot{\rho}}\left[\hat{\rho}^{c} - \rho_{m}S_{mH}\right]\ .
\end{equation}
Through (\ref{pcnad}) the dynamics of the total energy-density perturbations, described by Eq.~(\ref{dddeltak}), is coupled to $S_{mH}$. Explicitly,
\begin{equation}
\delta^{c\prime\prime} + \left[\frac{3}{2}-\frac{15}{2}\frac{p}{\rho}+ 3\frac{\dot{p}}{\dot{\rho}}\right]
\frac{\delta^{c\prime}}{a}
- \left[\frac{3}{2} + 12\frac{p}{\rho} - \frac{9}{2}\frac{p^{2}}{\rho^{2}} - 9\frac{\dot{p}}{\dot{\rho}}
 - \frac{k^{2}}{a^{2}H^{2}}\frac{\dot{p}}{\dot{\rho}}\right]\frac{\delta^{c}}{a^{2}}
 = \frac{k^{2}}{a^{2}H^{2}}\frac{\dot{p}}{\dot{\rho}}\frac{\rho_{m}}{\rho}\frac{S_{mH}}{a^{2}}
\ .
  \label{prprdeltaS}
\end{equation}
At high redshift, for $a\ll 1$ Eq. (\ref{prprdeltaS}) approaches the Einstein-de Sitter limit
\begin{equation}\label{EdS}
\delta^{c\prime\prime} + \frac{3}{2a}\delta^{c\prime} - \frac{3}{2a^{2}}\delta^{c} = 0
\qquad (a \ll 1) \ .
\end{equation}

To obtain an expression for the term on the right-hand side of Eq.~(\ref{ddSmHad}) we write
\begin{equation}\label{}
\hat{p}^{c_{_{H}}} = \hat{p} + \dot{p}v_{H}
= \hat{p} + \dot{p}v + \dot{p}\left(v_{H} - v\right)
\end{equation}
or
\begin{equation}\label{}
\hat{p}^{c_{_{H}}}
= \hat{p}^{c} + \dot{p}\left(v_{H} - v\right)\ .
\end{equation}
Now, for the difference $v - v_{H}$ we have
\begin{equation}\label{}
v - v_{H} = \frac{\rho_{m}}{\rho + p}\left(v_{m} - v_{H}\right)\ .
\end{equation}
From relation (\ref{dSmH1}) it follows that (in $k$ space)
\begin{equation}\label{}
v_{m} - v_{H} = \frac{a^{2}}{k^{2}}\dot{S}_{mH}\ .
\end{equation}
Then $\hat{p}^{c_{_{H}}}$ is written
\begin{equation}\label{}
\hat{p}^{c_{_{H}}}
= \hat{p}^{c} - \dot{p}\frac{\rho_{m}}{\rho + p}\frac{a^{2}}{k^{2}}\dot{S}_{mH}
\quad \Rightarrow\quad \hat{p}^{c_{_{H}}} = \hat{p}^{c} + 3H \frac{\dot{p}}{\dot{\rho}} \rho_{m}\frac{a^{2}}{k^{2}}\dot{S}_{mH}
\ .
\end{equation}
Here we introduce $\hat{p}^{c}$ from (\ref{pcnad}) to obtain
\begin{equation}\label{hatpcH}
\hat{p}^{c_{_{H}}} = \frac{\dot{p}}{\dot{\rho}}\left[\rho\delta^{c} - \rho_{m}
\left(S_{mH} - 3H \frac{a^{2}}{k^{2}}\dot{S}_{mH}\right)\right]\ .
\end{equation}
By relation (\ref{hatpcH}) the dynamics of $S_{mH}$ in Eq.~(\ref{ddSmH}) is coupled to the dynamics of $\delta^{c}$.
Explicitly,
\begin{equation}\label{prprS}
S^{\prime\prime}_{mH}
+ \left[\frac{3}{2} - 3 \frac{r}{1+\omega}\frac{\dot{p}}{\dot{\rho}} - \frac{3}{2}\frac{p}{\rho}\right]\frac{S^{\prime}_{mH}}{a}
+ \frac{r}{1+\omega}\frac{\dot{p}}{\dot{\rho}}\frac{k^{2}}{a^{2}H^{2}}\frac{S_{mH}}{a^{2}}
= \left(1+\frac{r}{1+\omega}\right)\frac{\dot{p}}{\dot{\rho}}\frac{k^{2}}{a^{2}H^{2}}\frac{\delta^{c}}{a^{2}}\ ,
\end{equation}
where we have to exclude the case $\omega = -1$. Below we shall show that $\omega = -1$ is only possible asymptotically and with $\frac{\dot{p}}{\dot{\rho}} = 0$.
For $a \ll 1$ this equation approaches
\begin{equation}\label{}
S^{\prime\prime}_{mH} + \frac{3}{2a}S^{\prime}_{mH} = 0 \qquad (a\ll 1)\ .
\end{equation}
There is one decaying solution and a solution $S_{mH} = $ const.
At high redshift, according to (\ref{asmall}), the DE behaves as dust as well and we may use approximately adiabatic initial conditions $S_{mH} \approx 0 $ for the coupled system (\ref{prprdeltaS}) and (\ref{prprS}).

Our interest is the matter-energy perturbation. To this purpose we decompose
the total energy-density perturbation $\delta^{c}$ according to
\begin{equation}\label{}
\delta^{c} = \frac{\rho_{m}}{\rho}\delta_{m}^{c} + \frac{\rho_{H}}{\rho}\delta_{H}^{c} \ .
\end{equation}
Combination with $S_{mH} = \delta_{m} - \frac{\delta_{H}}{1+\omega}$ leads to
\begin{equation}\label{deltamc}
\delta_{m}^{c}= \frac{1}{1 + \frac{\omega}{1+r}}\left[\delta^{c} + \frac{1+\omega}{1+r}S_{mH}\right]\ ,
\end{equation}
which describes the matter-energy perturbations as a combination of $\delta^{c}$ and $S_{mH}$. To obtain its dynamics one has to solve the coupled system of equations (\ref{prprdeltaS}) and (\ref{prprS}).

The result for the matter perturbations of our model can be compared with the behavior of matter perturbations in the $\Lambda$CDM model. The latter can be obtained as a limiting case from Eq.~(\ref{prprdeltaS}) for the total energy-density perturbations. Using there
$p = p_{\Lambda} = - \rho_{\Lambda} = \mathrm{constant}$
and
\begin{equation}\label{}
\rho = \rho_{M} + \rho_{\Lambda}\ , \qquad \frac{p}{\rho} = - \frac{\rho_{\Lambda}}{\rho_{\Lambda}+\rho_{M}}
= - \frac{1}{1 + r}\ , \qquad r = r_{0}a^{-3} \qquad (\Lambda \mathrm{CDM})\ ,
\end{equation}
Eq.~(\ref{prprdeltaS}) reduces to
\begin{equation}
\delta^{c\prime\prime} + \left[\frac{3}{2}-\frac{15}{2}\frac{p}{\rho}\right]\frac{\delta^{c\prime}}{a}
- \left[\frac{3}{2} + 12\frac{p}{\rho} - \frac{9}{2}\frac{p^{2}}{\rho^{2}}
\right]\frac{\delta^{c}}{a^{2}}
= 0 \qquad (\Lambda \mathrm{CDM})\ .
  \label{dddla}
\end{equation}
For the perturbations we have
\begin{equation}\label{}
    \hat{\rho}^{c} = \hat{\rho}_{m}^{c} \quad \Rightarrow\quad \delta^{c} = \frac{\hat{\rho^{c}}}{\rho} = \frac{\hat{\rho}_{m}^{c}}{\rho_{\Lambda}+\rho_{m}} = \delta_{m}^{c}\frac{r}{1+r}\qquad (\Lambda \mathrm{CDM})\ .
\end{equation}
Then, the dynamics of $\delta_{m}^{c}$ is governed by
\begin{equation}\label{}
\delta_{m}^{c\prime\prime} + \frac{3}{2}\,\frac{2+r}{1+r}\,\frac{\delta_{m}^{c\prime}}{a}
- \frac{3}{2}\,\frac{r}{1+r}\,\frac{\delta_{m}^{c}}{a^{2}} = 0\ ,  \qquad (\Lambda \mathrm{CDM})\ .
\end{equation}
While for $r\ll 1$ at high redshift the behavior of $\delta_{m}^{c}$ for our Ricci-type DE model is
indistinguishable from that of the $\Lambda$CDM model, the future evolution is different.
The ``growing" mode of the latter approaches $\delta_{m}^{c} = $ constant for $r \rightarrow 0$, independently
of the scale, whereas for Ricci-DE the behavior of the corresponding quantity depends on the scale
and decays and/or oscillates.
Moreover, as we shall demonstrate below, the matter perturbations generally exhibit
instabilities at finite values  of the scale factor. Before clarifying this issue, a
comment concerning the interpretation of the perturbation variable is in order here.
The matter density perturbation $\delta_{m}^{c}$ in relation (\ref{deltamc}) is defined with respect to the \textit{total} comoving
gauge. To obtain the matter density perturbation, comoving with the matter velocity,
$\delta_{m}^{c_{m}} = \delta_{m} + \frac{\dot{\rho}_{m}}{\rho_{m}}v_{m}$,
 we have to consider
\begin{equation}\label{}
\delta_{m}^{c} = \delta_{m} + \frac{\dot{\rho}_{m}}{\rho_{m}}v =  \delta_{m} + \frac{\dot{\rho}_{m}}{\rho_{m}}v_{m}
+\frac{\dot{\rho}_{m}}{\rho_{m}}\left(v-v_{m}\right) = \delta_{m}^{c_{m}}
+ \frac{\dot{\rho}_{m}}{\rho_{m}}\left(v-v_{m}\right)\ .
\end{equation}
Since
\begin{equation}\label{}
v-v_{m} = \frac{\rho_{H} + p_{H}}{\rho+p}\left(v_{H} - v_{m}\right)\ , \qquad v_{H} - v_{m}
= - \frac{a^{2}}{k^{2}}\dot{S}_{mH}\ ,
\end{equation}
the quantity of interest is
\begin{equation}\label{deltamcm}
\delta_{m}^{c_{m}} = \frac{1}{1 + \frac{\omega}{1+r}}\left[\delta^{c} + \frac{1+\omega}{1+r}S_{mH}\right]
- \frac{3}{1 + \frac{r}{1+\omega}}\frac{a^{2}H^{2}}{k^{2}}aS^{\prime}_{mH}\ .
\end{equation}
Obviously, $\delta_{m}^{c}$ and $\delta_{m}^{c_{m}}$ differ by the last term in (\ref{deltamcm}).
Because of the factor $\frac{a^{2}H^{2}}{k^{2}}$ (assuming again $\omega \neq -1$) one expects that on scales smaller than the Hubble scale
the differences between $\delta_{m}^{c}$ and $\delta_{m}^{c_{m}}$ are small.

\section{Issues of stability and a viable model}
\label{stab}

\begin{figure}[!t]
\begin{center}
\begin{minipage}[t]{0.50\linewidth}
\includegraphics[width=\linewidth]{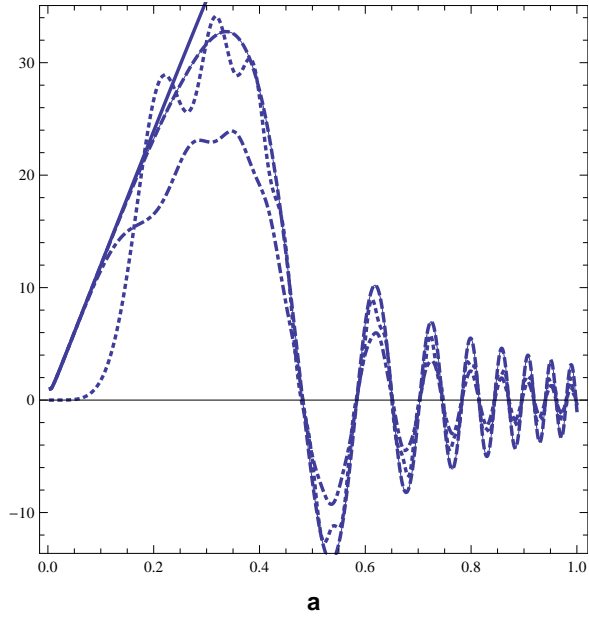}
\end{minipage} \hfill
\caption{Evolution of the perturbation quantities $S_{mH}$ (dotted line), $\delta^{c}$ (dash-dotted line),
$\delta^{c}_{m}$ (dashed line) and $\delta^{c_{_{m}}}_{m}$ (thin solid line)  for $\omega_{0} = -0.9$, $r_{0} = 0.4$ and $k= 0.1$. For comparison the $\Lambda$CDM result (thick solid line) is also included.}
\label{fig3}
\end{center}
\end{figure}

\begin{figure}[!t]
\begin{center}
\begin{minipage}[t]{0.50\linewidth}
\includegraphics[width=\linewidth]{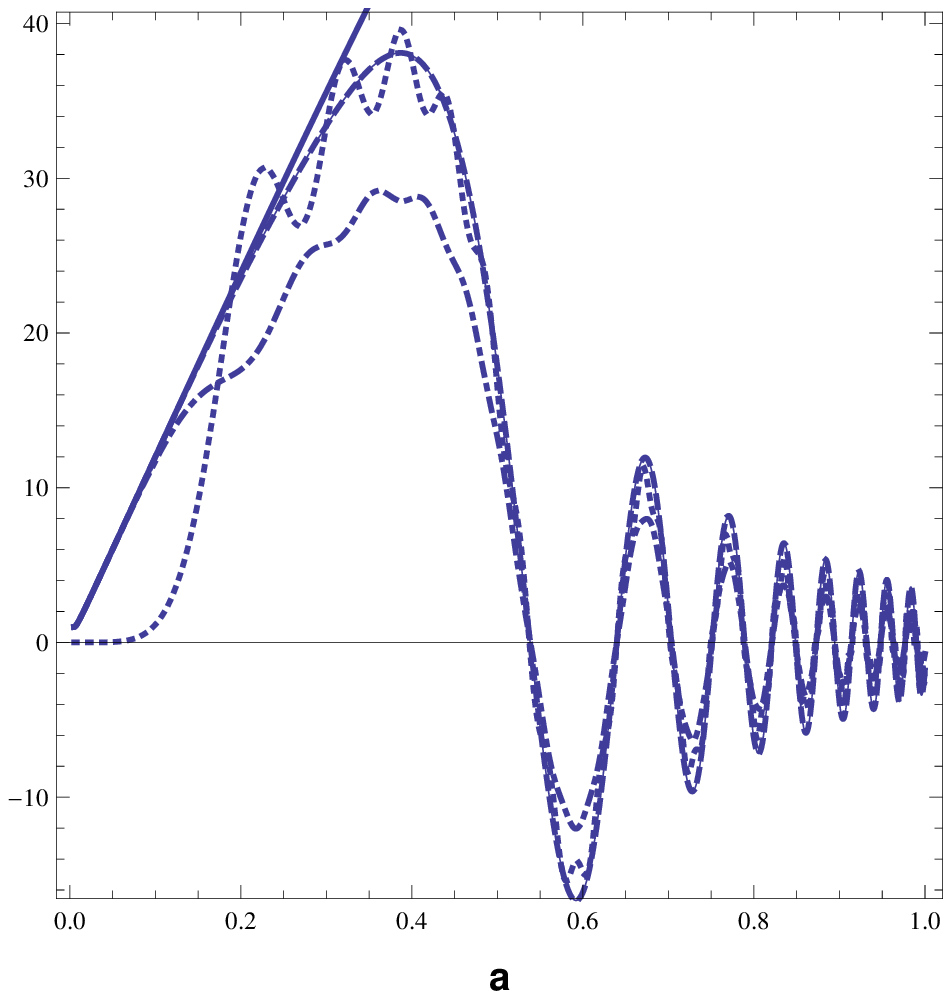}
\end{minipage} \hfill
\caption{The same quantities as in Fig. \ref{fig3} for $\omega_{0} = -0.9$, $r_{0} = 0.8$ and $k= 0.1$.}
\label{fig4}
\end{center}
\end{figure}

\begin{figure}[!t]
\begin{center}
\begin{minipage}[t]{0.50\linewidth}
\includegraphics[width=\linewidth]{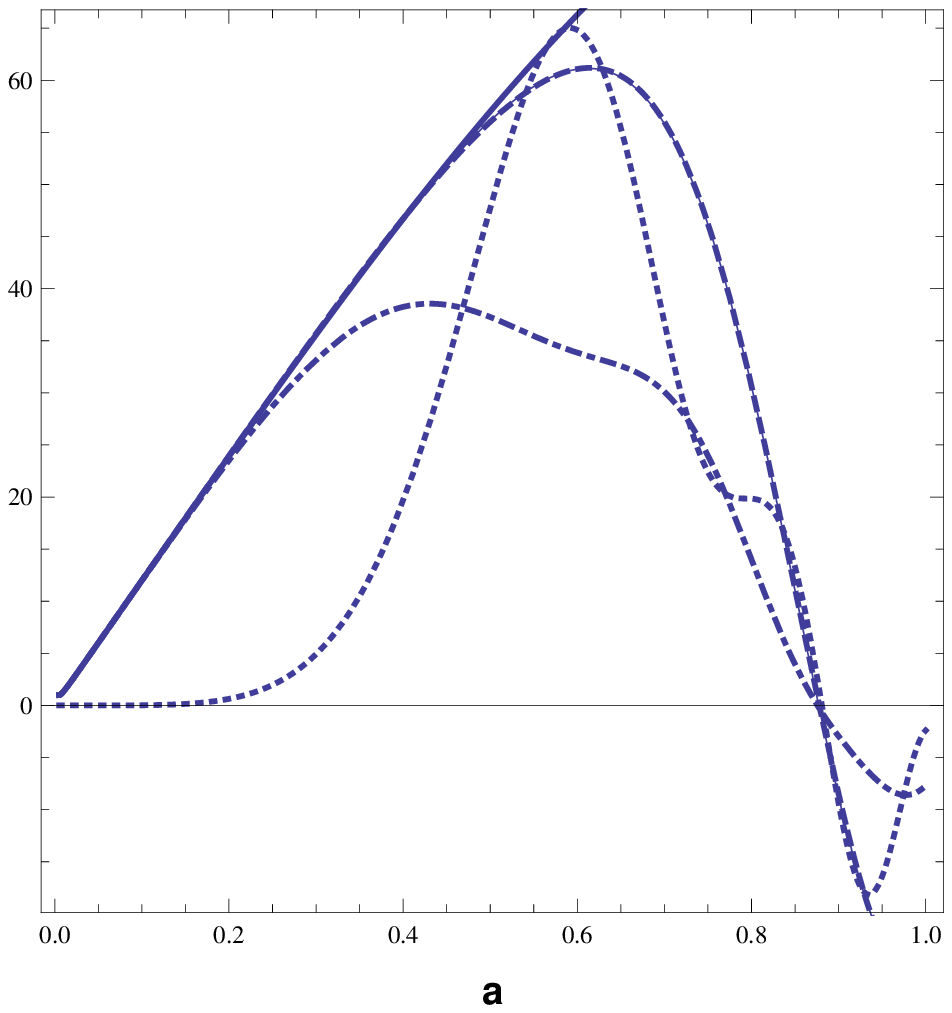}
\end{minipage} \hfill
\caption{The same quantities as in Fig. \ref{fig3} for $\omega_{0} = -0.9$, $r_{0} = 0.4$ and $k= 0.01$.}
\label{fig5}
\end{center}
\end{figure}

\begin{figure}[!t]
\begin{center}
\begin{minipage}[t]{0.50\linewidth}
\includegraphics[width=\linewidth]{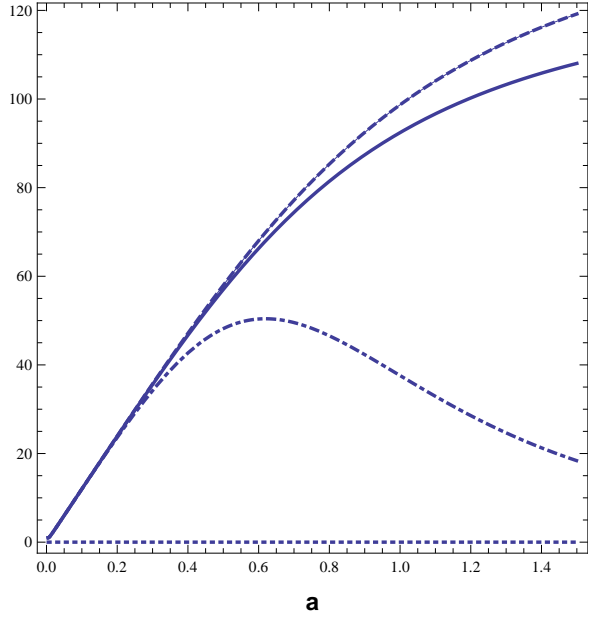}
\end{minipage} \hfill
\caption{Evolution of the perturbation quantities $S_{mH}$ (dotted line), $\delta^{c}$ (dash-dotted line),
$\delta^{c}_{m}$ (dashed line) and $\delta^{c_{_{m}}}_{m}$ (thin solid line) for $r_{0} = 0.4$ and $k= 0.1$ on the basis of (\ref{wsol+}), (\ref{r+}) and (\ref{dotp0}).
The $\Lambda$CDM result is represented by the thick solid line. The relative density perturbations $S_{mH}$ are negligible during the entire evolution. The results for $\delta^{c}_{m}$ and $\delta^{c_{_{m}}}_{m}$ are almost identical.}
\label{figura1}
\end{center}
\end{figure}

\begin{figure}[!t]
\begin{center}
\begin{minipage}[t]{0.50\linewidth}
\includegraphics[width=\linewidth]{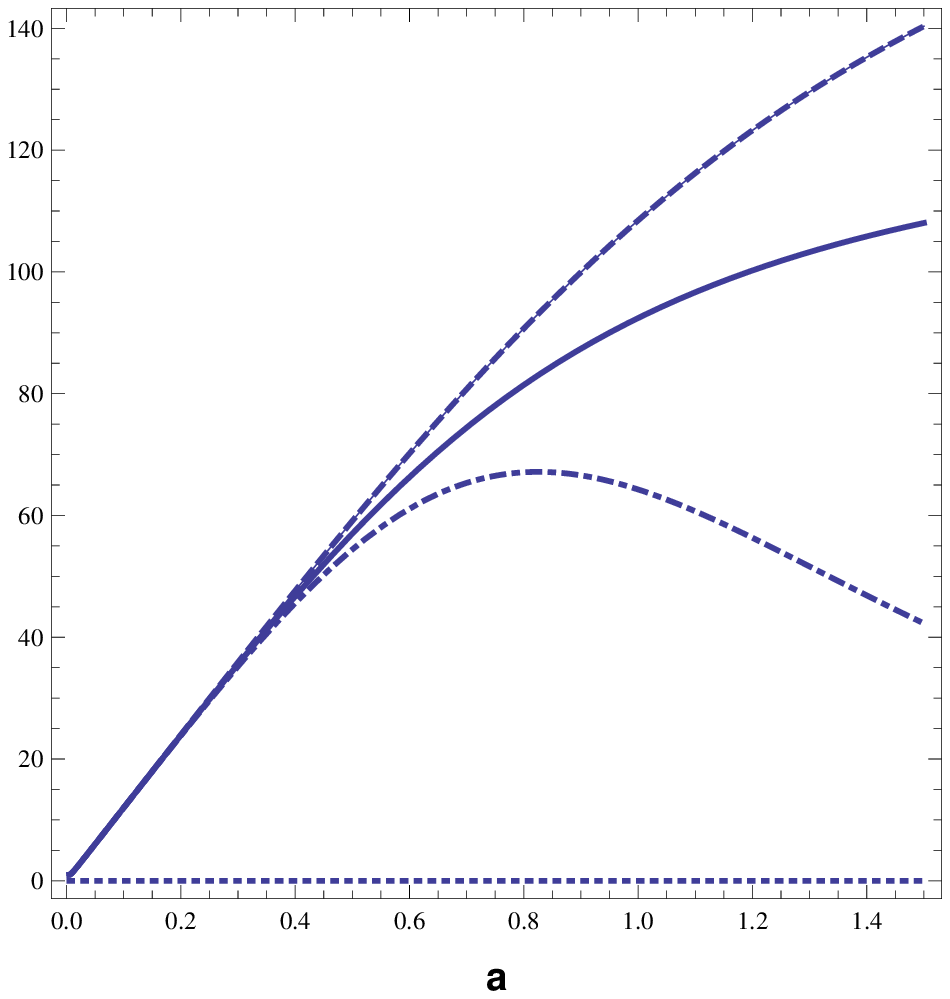}
\end{minipage} \hfill
\caption{The same quantities as in Fig. \ref{figura1} for $r_{0} = 0.8$ and $k= 0.1$.}
\label{figura2}
\end{center}
\end{figure}

\begin{figure}[!t]
\begin{center}
\begin{minipage}[t]{0.50\linewidth}
\includegraphics[width=\linewidth]{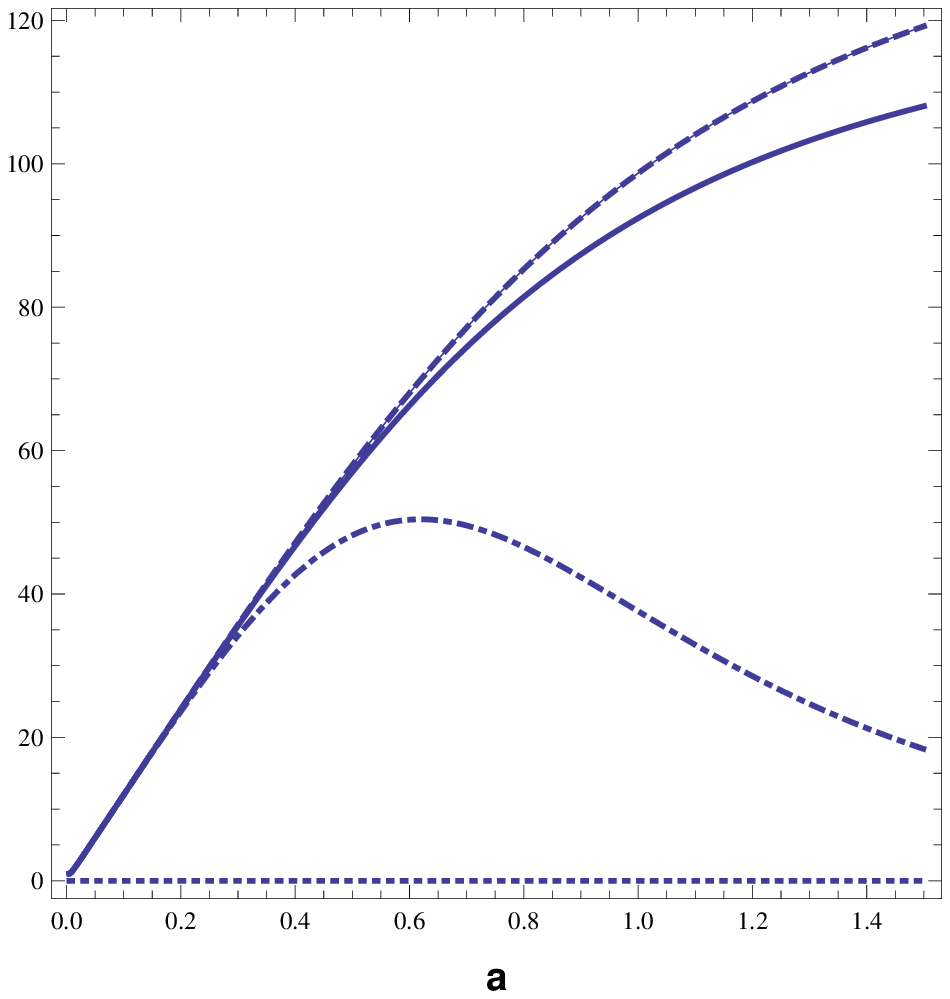}
\end{minipage} \hfill
\caption{The same quantities as in Fig. \ref{figura1} for $r_{0} = 0.4$ and $k= 0.01$.}
\label{figura3}
\end{center}
\end{figure}

The behavior of the quantities  $S_{mH}$,  $\delta^{c}$, $\delta_{m}^{c}$ and $\delta_{m}^{c_{_{m}}}$
together with $\delta_{m}^{c}$ of the $\Lambda$CDM model is visualized in Figs.~\ref{fig3} and \ref{fig5} for $\omega_{0} = - 0.9$ with  $k=0.1$ and $k=0.01$, respectively.
The parameters in Fig.~\ref{fig4} differ from those of Fig.~\ref{fig3} by a higher energy-density ratio.
This indicates a weak dependence on $r_{0}$.
The figures confirm that differences between $\delta_{m}^{c}$ and $\delta_{m}^{c_{m}}$ are indeed small
on the chosen scales.
In Figs. \ref{fig3} and \ref{fig4} there appear oscillations of all the perturbation quantities very close to the present time. They seem to be similar to those known from (generalized) Chaplygin-gas models which
have jeopardized these models (cf. \cite{Sandvik,w1} for a discussion of the matter power spectrum in the context of Chaplygin-gas models).
A still more serious drawback is the existence of instabilities at future values $a > 1$ (for $\omega_{0} > -1$) of the scale factor, related to a crossing of the phantom divide $\omega = -1$. Instabilities occur if the denominator $1 + \omega$ in (\ref{prprS}) vanishes, i.e., if $\omega$ approaches $-1$.
The condition of $1 + \omega = 0$ is the vanishing of the numerator in (\ref{wsol}),
\begin{equation}\label{}
r_{0} = \omega_{0}\left(3 - \left(r_{0} - 3\omega_{0}\right)\right)a_{i}^{r_{0} - 3\omega_{0}}\ ,
\end{equation}
which determines the value $a_{i}$ of the scale factor at which the instability occurs. Solving for $a_{i}$
yields
\begin{equation}\label{}
a_{i}^{r_{0} - 3\omega_{0}} = \frac{r_{0}}{\omega_{0}\left(3 - \left(r_{0} - 3\omega_{0}\right)\right)}
= \frac{r_{0}}{\omega_{0}\left(3\left(1+\omega_{0}\right) - r_{0}\right)} \ .
\end{equation}
We may now consider separately the cases $\omega_{0} > -1$ and $\omega_{0} < -1$.
Assuming $\omega_{0} = - 1 + \mu$ we have ($\mu \neq \frac{r_{0}}{3}$ and $\mu \neq 1$)
\begin{equation}\label{}
a_{i}^{r_{0} - 3\omega_{0}} = \frac{r_{0}}{\left(r_{0} - 3\mu\right)\left(1 - \mu\right)}\ .
\end{equation}
For $\mu > 0$  we find $a_{i} > 1$, i.e., the instability sets in at a finite value of the scale factor
in the future. This corresponds to the situation of Figs. \ref{fig3}, \ref{fig4} and \ref{fig5}, where the
instability sets in for values of the order of $a\approx 1.5$.
For $\mu < 0$, i.e. for a phantom equation of state, there appears an instability in the past
at $a_{i} < 1$. Since such kind of instability has not been observed, a present phantom equation of state
is definitely excluded in the context of our model. The limit between the two regimes is just $\mu = 0$ where
we have $a_{i} = 1$, i.e., instabilities at the present epoch.
The only case without instabilities at finite values of the scale factor is a fixed relation
$r_{0} - 3\omega_{0} = 3$ between the initially independent values of $r_{0}$ and $\omega_{0}$.
Since $r_{0} > 0$ necessarily, this implies $\omega_{0} > - 1$.
Consequently, the only physically acceptable case is
\begin{equation}\label{acceptable}
\omega_{0} = - 1 + \frac{r_{0}}{3} \quad \Leftrightarrow\quad r_{0} = 3 \left(1 + \omega_{0}\right)
\quad \Leftrightarrow\quad \Omega_{m0} = 3 \frac{1+\omega_{0}}{1+3\left(1+\omega_{0}\right)}\ .
\end{equation}
The parameters $\omega_{0}$ and $r_{0}$ are necessarily related to each other and cannot be chosen
independently.
In a sense, $r_{0}$ quantifies the deviation of $\omega_{0}$ from $\omega_{0} = - 1$. Under this condition we have $c^2 = \frac{1}{2}$. This is exactly the result found by Karwan and Thitapura in their study of instabilities through nonadiabatic perturbations in a system of matter and Ricci DE \cite{karwan}.
The solution (\ref{wsol}) then simplifies to
\begin{equation}\label{wsol+}
\omega = \frac{\omega_{0}}{\left(1+\omega_{0}\right)a^{-3} - \omega_{0}}
\end{equation}
and the solution (\ref{r}) becomes
\begin{equation}\label{r+}
r = 3 \frac{r_{0}}{\left(3 - r_{0}\right)a^{3} +r_{0}} \ .
\end{equation}
Combining the relations (\ref{wsol+}) and (\ref{r+}) has the important consequence
\begin{equation}\label{dotp0}
\frac{r}{1 + \omega} =3\qquad \Rightarrow\qquad \frac{\dot{p}}{\dot{\rho}} = 0\ .
\end{equation}
This makes all the coupling terms (and some others) in the coupled system (\ref{prprdeltaS}) and (\ref{prprS}) vanish.
Also the pressure perturbations $\hat{p}^{c}$ in (\ref{pcnad}) vanish.
There remain neither oscillations nor instabilities.
Equation (\ref{prprdeltaS}) reduces to the corresponding Eq. (\ref{dddla})  of the $\Lambda$CDM model.
However, the coefficients in both equations describe a different background dynamics which leads to different growth rates for the matter perturbations.
The relative energy-density perturbation remain negligible during the entire evolution.
From relations (\ref{wsol+}) and (\ref{r+}) we obtain that at high redshift
\begin{equation}\label{}
\omega \rightarrow 0 \quad \mathrm{and }\quad r \rightarrow 3    \qquad (a\ll 1)\ ,
\end{equation}
respectively, while in the far-future
\begin{equation}\label{}
\omega \rightarrow -1\quad \mathrm{and }\quad r \rightarrow   0  \qquad (a\gg 1)\ ,
\end{equation}
respectively, are valid.
The Hubble rate is given by
\begin{equation}\label{Hstab}
\frac{H^{2}}{H_{0}^{2}} = \Omega_{m0}a^{-3} + 1 + \frac{1}{3}\Omega_{m0}\left(a^{-3} - 4\right)\ .
\end{equation}
Notice that we have the same number of free parameters as in the $\Lambda$CDM model, but there is no
$\Lambda$CDM limit of (\ref{Hstab}).
The behavior of the perturbation quantities on the basis of (\ref{wsol+}), (\ref{r+}) and (\ref{dotp0}) is visualized in Figs. \ref{figura1}, \ref{figura2} and
\ref{figura3}. These figures confirm that for the chosen configuration there are neither oscillations nor instabilities. From this point of view the model appears acceptable. But
the observationally preferred (from the background tests in Sec. \ref{observations}) values of $\omega_{0}$ and $\Omega_{m0}$ are only marginally consistent with the instability-avoiding combination of these quantities.
The value $r_{0} = 0.4$, equivalent to $\Omega_{m0} = 0.27$, corresponds, according to relation (\ref{acceptable}), to a present value of the EoS parameter of $\omega_{0} = - 0.87$ which  differs
from the preferred background value $\omega_{0} = - 0.987$.
The situation, which is visualized in Fig.~\ref{fig2}, could be improved by a very low present matter content although this is not supported by the recent results of the Planck satellite \cite{planck}. On the other hand, both the results from SNIa-observations and those of the Planck satellite rely crucially on the
$\Lambda$CDM model which, as we have pointed out, does not follow as a limiting case from our Ricci-DE model.
Moreover, we have neglected here the baryon component.



\section{Conclusions}
\label{conclusions}

By its definition, noninteracting Ricci-type DE is characterized by a necessarily time-dependent EoS parameter. This makes it an observationally testable alternative to the $\Lambda$CDM model. It establishes a
a relationship between this EoS parameter and the matter content of the Universe. Ricci-DE behaves almost as dust at high redshift. The ratio of the energy densities of DM and DE varies considerably less than for the
 $\Lambda$CDM model. Since the time of radiation decoupling it has changed by about one order of magnitude
 compared with roughly nine orders of magnitude for the  $\Lambda$CDM model. This amounts to a remarkable alleviation of the coincidence problem. Our statistical analysis, based on recent observational data from SNIa, BAO and $H(z)$, results in a preferred value of $c^{2}\approx 0.46$ for the Ricci-DE parameter which confirms earlier studies in the literature \cite{gao}.
 Within a gauge-invariant analysis we calculated the matter perturbations as a combination of the total energy perturbations of the cosmic medium and the relative perturbations of the components and clarified the relation to the matter perturbations of the $\Lambda$CDM model.
The results coincide with those of the $\Lambda$CDM model until redshifts of the order of $z \approx 4$ . For $z< 1.5$   the differences are substantial.

 We demonstrated that the perturbation dynamics suffers from instabilities that exclude a present phantom-type equation of state.
 For values $\omega_{0} > -1$ the theory predicts instabilities at finite future values of the scale factor.
 It is only for a specific relation between the values $\Omega_{m0}$ of the present matter density and the present EoS parameter $\omega_{0}$ that the dynamics remains stable for any finite scale-factor value.
 This relation corresponds to a Ricci-DE parameter $c^{2}= 0.5$ \cite{karwan}.
 The cosmological evolution is then governed by dynamical DE with time-varying EoS and vanishing pressure perturbations. The number of independent parameters of this model reduces to that of the $\Lambda$CDM model.
 The basic equation for the perturbation dynamics formally coincides with its $\Lambda$CDM counterpart as well.  But due to the differences in the background dynamics the growth rates of the matter perturbations are different.
 We conclude that holographic Ricci DE represents a theoretically appealing scenario which does not need additional parameters except $H_{0}$ and $\Omega_{m0}$. Despite of its  attractive features,
there remains a certain tension between the instability-avoiding theoretical values of $\Omega_{m0}$ and $\omega_{0}$  and those preferred by the analysis of the homogeneous and isotropic background dynamics.

\acknowledgments{The authors would like to thank Oliver Piattella and Giuseppe Dito  for helpful discussions. This work was supported by the ``Comisi\'{o}n
Nacional de Ciencias y Tecnolog\'{\i}a" (Chile) through the
FONDECYT Grants No. 1110230 and  No. 1130628 (R.H. and S.d.C).
J.C.F and W.Z acknowledge support by ``FONDECYT-Concurso incentivo a la
Cooperaci\'{o}n Internacional" No. 1130628 as well as by CNPq (Brazil) and FAPES (Brazil).}


\begin{thebibliography}{99}



\bibitem{SN}A. G. Riess et al., Astron. J. \textbf{116}, 1009
(1998); S. J. Perlmutter et al., Astrophys. J.
\textbf{517}, 565 (1999); A. G. Riess et al., Astrophys. J. \textbf{607}, 665 (2004);
P. Astier et al., Astron. Astrophys. \textbf{447}, 31 (2006).

\bibitem{wmap9} G. Hinshaw et al., arXiv:1212.5226.

\bibitem{problems} S. Weinberg, Rev. Mod. Phys. \textbf{61}, 1 (1989);
 E.J. Copeland, M. Sami, S. Tsujikawa, Int. J. Mod. Phys.  D \textbf{15}, 1753 (2006).

\bibitem{cohen}
A. G. Cohen, D.B. Kaplan and A.E. Nelson, Phys. Rev. Lett.
\textbf{82}, 4971 (1999).

\bibitem{li}
M. Li, Phys. Lett. B \textbf{603}, 1 (2004).

\bibitem{Hsu}
S. D. H. Hsu, Phys. Lett. B \textbf{594}, 13 (2004).

\bibitem{DW}
D. Pav\'{o}n and W. Zimdahl, Phys. Lett. B \textbf{628}, 206
(2005).

\bibitem{HDE}
 W. Zimdahl and D. Pav\'{on}, Class. Quantum Grav. \textbf{24}, 5641 (2007); W. Zimdahl, IJMPD  \textbf{17}, 651 (2008).

 \bibitem{futureEH} Hsien-Chung Kao, Wo-Lung Lee and Feng-Li Lin, Phys. Rev. D \textbf{71}, 123518 (2005);
K. Enquist, S. Hannestad and M.S. Sloth, JCAP \textbf{0202}, (2005) 004;
Yungui Gong, Bin Wang and Yuan-Zhong Zhang, Phys. Rev. D \textbf{72}, 043510 (2005);
Xin Zhang and Feng-Quan Wu, Phys. Rev. D \textbf{72}, 043524 (2005); Phys. Rev. D \textbf{76}, 023502 (2007);
Qiang Wu, Yungui Gong, Anzhong Wang andJ.S. Alcaniz, Phys. Lett. B \textbf{659}, 34 (2008);
Shao-Feng Wu, Peng-Ming Zhang, and Guo-Hong Yang, Class. Quantum Grav. \textbf{26} (2009) 055020;
Yinzhe Ma, Yan Gong and Xuelei Chen, arXiv:0901.1215.


\bibitem{brustein}
R. Brustein and G. Veneziano, Phys. Rev. Lett. \textbf{84},
5695 (2000).


\bibitem{gao} C. Gao, F. Q. Wu, X. Chen and Y. G. Shen, Phys. Rev. D
\textbf{79}, 043511 (2009).

\bibitem{cai}
R-G. Cai, B. Hu and Y. Zhang, Commun. Theor. Phys. \textbf{51}, 954 (2009).


\bibitem{xuliluchang} Lixin Xu, Wenbo Li, Jianbo Lu, and Baorong Chang, Mod.Phys.Lett.A \textbf{24}, 1355 (2009),  arXiv:0810.4730.

\bibitem{xuwang} Lixin Xu and Yuting Wang, JCAP \textbf{1006}, 002 (2010);
arXiv:1006.0296.


\bibitem{xinzhang}
Xin Zhang, Phys. Rev. D\textbf{79}, 103509 (2009); arXiv:0901.2262.

\bibitem{rong}
Rong-Jia Yang, Zong-Hong Zhu and Fengquan Wu, Int. J. Mod. Phys. A \textbf{26}, 317 (2011); arXiv:1101.4797.

\bibitem{fengli} Chao-Jun Feng and Xin-Zhou Li, Phys. Lett. B \textbf{680}, 184 (2009); arXiv:0904.2972.

\bibitem{karwan} K. Karwan and T. Thitapura, JCAP \textbf{1201}, 017 (2012); arXiv:1110.2451.

\bibitem{yuting} Yuting Wang, Lixin Xu and Yuanxing Gui, Phys. Rev. D \textbf{84}, 063513 (2011); arXiv:1110.4401.

    \bibitem{fengli} Chao-Jun Feng and Xin-Zhou Li, Phys.Lett. \textbf{B680}, 355 (2009), arXiv:0905.0527.

\bibitem{yi}
Yi Zhang and Hui Li, arXiv:1003.2788.

\bibitem{zhang3} Jingfei Zhang, Li Zhang, and Xin Zhang, Phys.Lett. \textbf{B691}, 11 (2010), arXiv:1006.1738.

\bibitem{ivdi}
I. Dur\'{a}n and D. Pav\'{o}n, Phys. Rev. D \textbf{83}, 023504 (2011); arXiv:1012.2986.


\bibitem{luis11} L.P. Chimento and M.G. Richarte, Phys. Rev. D \textbf{84}, 123507 (2011), arXiv:1107.4816.


\bibitem{tian-fu} Tian-Fu Fu, Jing-Fei Zhang, Jin-Qian Chen and Xin Zhang, Eur. Phys. J. C \textbf{72}, 1932 (2012); arXiv:1112.2350.

    \bibitem{jamil1} A. Pasqua, A. Khodam-Mohammadi, M. Jamil and R. Myrzakulov,
    Astrophys.Space Sci. \textbf{340}, 199 (2012), arXiv:1112.6381.

\bibitem{zhenhui} Zhenhui Zhang, Song Li,  Xiao-Dong Li, Xin Zhang and Miao Li, JCAP \textbf{1206}, 009 (2012);  arXiv:1204.6135.




\bibitem{luis} L.P. Chimento and M.G. Richarte, Phys. Rev. D \textbf{85}, 127301 (2012); arXiv:1207.1492.

\bibitem{jamil2} A. Pasqua, M. Jamil, R. Myrzakulov and B. Majeed, Phys. Scr. \textbf{86}, 045004 (2012), arXiv:1211.0902

\bibitem{luis13} L.P. Chimento, M. Forte  and M.G. Richarte, Eur.Phys.J.C \textbf{73}, 2285 (2013), arXiv:1301.2737.

\bibitem{broda} B. Broda, IJMPD \textbf{21}, 1250053 (2012);  arXiv:1111.5785.

\bibitem{granda} L.N. Granda and A. Oliveros, Phys.Lett. \textbf{B669}, 275 (2008), arXiv:0810.3149.

\bibitem{statef} Chao-Jun Feng, Phys.Lett. \textbf{B670}, 231 (2008), arXiv:0809.2502.

\bibitem{fengzhang} Chao-Jun Feng and Xin Zhang, Phys.Lett. \textbf{B680}, 399 (2009), arXiv:0904.0045.


\bibitem{xululi} Lixin Xu, Jianbo Lu, and Wenbo Li, Eur.Phys.J.C \textbf{64}, 89 (2009), arXiv:0906.0210.

\bibitem{granda2} L.N. Granda, W. Cardonay and A. Oliveros, arXiv:0910.0778.

\bibitem{suwa} M. Suwa and T. Nihei, Phys.Rev.D \textbf{81}, 023519 (2010), arXiv:0911.4810.

\bibitem{feiyu} Fei Yu, Jingfei Zhang, Jianbo Lu, Wei Wang and Yuanxing Gui, Phys.Lett. \textbf{B688}, 263 (2010), arXiv:1004.2092.

\bibitem{yuzhang}  Fei Yu and Jing-Fei Zhang,  Commun.Theor.Phys. \textbf{59}, 243 (2013), arXiv:1305.2792.



\bibitem{SRJW} S. del Campo, J.C. Fabris, R. Herrera and W. Zimdahl,
Phys. Rev. D \textbf{83}, 123006 (2011).



\bibitem{jimenez} R. Jimenez and A. Loeb, Astrophys.J. \textbf{573}, 37 (2002); 37, arXiv:astro-ph/0106145.
\bibitem{verde} J. Simon, L. Verde and R. Jimenez, Phys. Rev. D {\bf71}, 123001 (2005).
\bibitem{stern} D. Stern, R. Jimenez, L. Verde, M. Kamionkowski and S. A. Stanford,
JCAP {\bf 1002}, 008 (2010).
\bibitem{verdebis} R. Jimenez, L. Verde, T. Treu and D. Stern, Astrophys. J. {\bf 593}, 622 (2003).
\bibitem{ma} T-J. Zhang, C. Ma and T. Lan, Adv. in Astron. {\bf 2010}, 184284 (2010).
\bibitem{mabis} C. Ma and T-J. Zhang, Astrophys. J. {\bf 730}, 74 (2011).
\bibitem{moresco} M. Moresco, A. Cimatti, R. Jimenez et al. JCAP \textbf{1208}, 006 (2012, arXiv:1201.3609.
\bibitem{ratra} O. Farooq, D. Mania and B. Ratra, {\it Hubble parameter measurement constraints on dark
energy}, arXiv:1211.4253.
\bibitem{ioav}
B.L. Lago, M.O. Calv\~{a}o, S.E. Jor\'{a}s, R.R.R. Reis, I. Waga and R. Giostri, {\it Type Ia supernova parameter estimation: a comparison of two approaches using current datasets}, arXiv:1104.2874.
\bibitem{union} R. Amanullah, et al. Astrophys. J. {\bf 716}, 712( 2010).
\bibitem{eise} D.L. Eisenstein et al. (SDSS) Astrophys. J. {\bf 633}, 560 (2005).
\bibitem{blake} C. Blake et all, {\it The WiggleZ Dark Energy Survey: mapping the distance-redshift relation with baryon acoustic oscillations}, arXiv:1108.2635.
\bibitem{Sandvik} H.B. Sandvik, M. Tegmark, M. Zaldariaga and I. Waga,
Phys. Rev. D \textbf{69}, 123524 (2004).

\bibitem{w1} W.S. Hip\'olito-Ricaldi, H.E.S. Velten and W. Zimdahl, Phys. Rev. D {\bf82}, 063507 (2010).
\bibitem{planck} Planck Collaboration:arXiv:1303.5076.
\end{thebibliography}
\end{document}